\documentclass[12pt,letterpaper]{article}
\pdfoutput=1

\usepackage{amsmath,amssymb,calc, amsthm,bbm, epsfig,psfrag, mathtools,comment}

\usepackage{dsfont}
\usepackage{mathrsfs}
\usepackage{graphicx, enumerate, enumitem}
\usepackage{float}

\usepackage[numbers,sort&compress]{natbib}

\usepackage{textcomp}
\usepackage{wasysym}
\usepackage{tabularx}
\usepackage{multicol}
\usepackage{longtable,booktabs,colortbl}

\usepackage[vcentermath]{youngtab}
\usepackage{cancel}

\usepackage[pdftex,table,svgnames,hyperref]{xcolor}

\definecolor{darkred}{rgb}{0.5 0 0}
\definecolor{darkgreen}{rgb}{0.5 .5 0}
\definecolor{darkblue}{rgb}{0 0 .5}

\usepackage{mathematica}
\usepackage{lieart}

\usepackage{tikz}
\usepackage{physics}
\usepackage{tikz-cd}
\usepackage{tensor}
\newcommand{\ul}{\underline}

\usepackage[utf8]{inputenc} 
\definecolor{azure}{rgb}{0.0, 0.5, 1.0}
\definecolor{darkblue}{rgb}{0.15,0.35,0.7}
\definecolor{reddish}{rgb}{0.65, 0.2, 0.2}
\definecolor{brandeisblue}{rgb}{0.0, 0.44, 1.0}
\definecolor{ceruleanblue}{rgb}{0.16, 0.32, 0.75}
\definecolor{indigo(dye)}{rgb}{0.0, 0.25, 0.42}
\usepackage[linktocpage=true]{hyperref}
\hypersetup{
colorlinks=true,
citecolor=ceruleanblue,
linkcolor=ceruleanblue,
urlcolor=ceruleanblue,
pdfauthor={},
pdftitle={},
pdfsubject={}
}

\newcommand{\overbar}[1]{\mkern 1.5mu\overline{\mkern-1.5mu#1\mkern-1.5mu}\mkern 1.5mu}

\newcommand{\TT}{T\overbar{T}}

\DeclareFontEncoding{LS1}{}{}
\DeclareFontSubstitution{LS1}{stix}{m}{n}
\DeclareSymbolFont{stixsymbols}{LS1}{stixscr}{m}{n}
\SetSymbolFont{stixsymbols}{bold}{LS1}{stixscr}{b}{n}
\DeclareMathSymbol{\kay}{\mathalpha}{stixsymbols}{"6B}
\DeclareMathSymbol{\hay}{\mathalpha}{stixsymbols}{"68}

\makeatletter
\renewcommand\section{\@startsection {section}{1}{\z@}%
                               {-3.5ex \@plus -1ex \@minus -.2ex}
                               {2.3ex \@plus.2ex}%
                               {\normalfont\large\bfseries}}
\renewcommand\subsection{\@startsection{subsection}{2}{\z@}%
                                 {-3.25ex\@plus -1ex \@minus -.2ex}%
                                 {1.5ex \@plus .2ex}%
                                 {\normalfont\bfseries}}
\makeatother

\makeatletter
\newcommand*\bigcdot{\mathpalette\bigcdot@{.5}}
\newcommand*\bigcdot@[2]{\mathbin{\vcenter{\hbox{\scalebox{#2}{$\m@th#1\bullet$}}}}}
\makeatother

\newfont{\goth}{ygoth.tfm scaled 1200}                   

\numberwithin{equation}{section}

\allowdisplaybreaks

\begin{document}
\begin{titlepage}
\begin{flushright}
\today
\end{flushright}
\vspace{5mm}

\begin{center}
{\Large \bf 
Machine Learning Invariants of Tensors}
\end{center}

\begin{center}

{\bf
Athithan Elamaran${}^{a}$,
Christian Ferko${}^{b, c}$,
Sterling Scarlett${}^{d}$
} \\
\vspace{5mm}

\footnotesize{
${}^{a}$
{\it 
Foothill High School, Pleasanton, CA 94588, USA
}
 \\~\\
${}^{b}$
{\it 
Department of Physics, Northeastern University, Boston, MA 02115, USA
}
 \\~\\
${}^{c}$
{\it 
The NSF Institute for Artificial Intelligence
and Fundamental Interactions
}
  \\~\\
${}^{d}$
{\it 
Department of Physics, Boston University, Boston, MA 02215, USA
}
}
\vspace{4mm}
~\\
\texttt{athithan.elamaran@gmail.com,
c.ferko@northeastern.edu,
sjscar@bu.edu
}\\
\vspace{2mm}

\end{center}

\begin{abstract}
\baselineskip=14pt

\noindent We propose a data-driven approach to identifying the functionally independent invariants that can be constructed from a tensor with a given symmetry structure. Our algorithm proceeds by first enumerating graphs, or tensor networks, that represent inequivalent contractions of a product of tensors, computing instances of these scalars using randomly generated data, and then seeking linear relations between invariants using numerical linear algebra. Such relations yield syzygies, or functional dependencies relating different invariants. We apply this approach in an extended case study of the independent invariants that can be constructed from an antisymmetric $3$-form $H_{\mu \nu \rho}$ in six dimensions, finding five independent invariants. This result confirms that the most general Lagrangian for such a $3$-form, which depends on $H_{\mu \nu \rho}$ but not its derivatives, is an arbitrary function of five variables, and we give explicit formulas relating other invariants to the five independent scalars in this generating set.

\end{abstract}
\vspace{5mm}

\vfill
\end{titlepage}


\renewcommand{\thefootnote}{\arabic{footnote}}
\setcounter{footnote}{0}

\tableofcontents{}
\vspace{0.5cm}
\hrule \bigskip

\section{Introduction}\label{sec:intro}

Many problems in mathematics and physics concern the construction and classification of \emph{invariants}: given one or more objects which transform under the action of a Lie group, what are the independent quantities that can be built from these objects and which are scalars under the group action? This problem arises, for instance, in the construction of Lagrangians for fields that transform under the Lorentz group. As a simple example, suppose that we wish to write down the most general Lagrangian in $4d$ which involves a field strength $F_{\mu \nu}$ but not its derivatives. There are two functionally independent invariants associated with this field strength, which we can choose to parameterize either via the traces
\begin{align}\label{x1_and_x2}
    x_1 = F_{\mu \nu} F^{\nu \mu} = \tr ( F^2 ) \, , \qquad x_2 = F_{\mu \nu} F^{\nu \rho} F_{\rho \sigma} F^{\sigma \mu} = \tr ( F^4 ) \, ,
\end{align}
or in terms of the Hodge dual $\widetilde{F}^{\mu \nu} = \frac{1}{2} \epsilon^{\mu \nu \rho \sigma} F_{\rho \sigma}$ as
\begin{align}\label{S_and_P}
    S = - \frac{1}{4} F_{\mu \nu} F^{\mu \nu} \, , \qquad P = - \frac{1}{4} F_{\mu \nu} \widetilde{F}^{\mu \nu} \, ,
\end{align}
or, by converting to spinor indices using the Weyl matrices $\left( \sigma^\mu \right)_{\alpha \dot{\alpha}}$ and $\left( \tilde{\sigma}^\nu \right)^{\dot{\alpha} \alpha}$ of $\mathrm{SL} ( 2 , \mathbb{C} )$ as $\tensor{F}{_\alpha^\beta} = - \frac{1}{4} \left( \sigma^\mu \right)_{\alpha \dot{\beta}} \left( \tilde{\sigma}^\nu \right)^{\dot{\beta} \beta} F_{\mu \nu}$ and $\tensor{\overbar{F}}{_{\dot{\alpha}}^{\dot{\beta}}} = \frac{1}{4} \left( \tilde{\sigma}^\mu \right)^{\dot{\beta} \beta} \left( \sigma^\nu \right)_{\beta \dot{\alpha}} F_{\mu \nu}$, as
\begin{align}\label{spinor_variables}
    \varphi = F^{\alpha \beta} F_{\alpha \beta} \, , \qquad \overbar{\varphi} = \overbar{F}_{\dot{\alpha} \dot{\beta}} \overbar{F}^{\dot{\alpha} \dot{\beta}} \, .
\end{align}
The most general theory in this class is therefore described by a Lagrangian which is a function of two variables, $\mathcal{L} ( x_1, x_2 )$ or $\mathcal{L} ( S, P )$ or $\mathcal{L} ( \varphi , \overbar{\varphi} )$. In this case, it is trivial to see that there are no other independent invariants: in components, $F_{\mu \nu}$ is a real antisymmetric $4 \times 4$ matrix, which has four purely imaginary eigenvalues occurring in complex conjugate pairs, and therefore contains only two pieces of Lorentz-invariant data. Parameterizing the most general Lagrangian for this field content is a useful starting point for studying other properties of theories of non-linear electrodynamics; one can impose constraints on the Lagrangian $\mathcal{L}$ which ensure other properties such as electric-magnetic duality invariance \cite{BialynickiBirula:1984tx,Gibbons:1995cv,Gaillard:1997rt,Gaillard:1997zr} or causality \cite{Schellstede:2016zue,Russo:2024llm}, study deformations of a seed theory driven by operators such as those built from the energy-momentum tensor \cite{Conti:2018jho,Babaei-Aghbolagh:2022uij,Ferko:2023ruw,Ferko:2023wyi}, and so on.

For objects that transform in more complicated representations of a group, or when considering a larger collection of such objects, the classification of singlets under the group action can become much more involved. This general problem belongs to the mathematical subject of \emph{invariant theory}; see, for instance, \cite{derksen2002computational} for an introduction. Other natural contexts in which invariant-theoretic questions arise include:
\begin{enumerate}[label = (\roman*)]

    \item\label{stringy} The organization of stringy corrections to effective actions. For instance, suppose we focus on the purely gravitational sector of the action for type IIA or type IIB supergravity. The $\alpha'$ expansion involves a series of terms that are Lorentz scalars constructed from the Riemann tensor $R_{\mu \nu \rho \sigma}$ in ten dimensions. What are the independent scalars that can be constructed from $R_{\mu \nu \rho \sigma}$, and given a new invariant which is known to be functionally dependent on those in some basis set, how can we express the new invariant in terms of the independent ones?

    \item\label{quantum} The classification of inequivalent classes of quantum states in finite-dimensional Hilbert spaces. Suppose that we are given a state $\ket{\psi} \in \mathcal{H}_1 \otimes \ldots \otimes \mathcal{H}_n$. Consider the action of the group of local unitary (LU) transformations on $\ket{\psi}$, or the more general group of \emph{stochastic local operations and classical communication} (SLOCC). What are the independent numerical quantities that can be constructed from the expansion coefficients of $\ket{\psi}$, in some basis for each of the $\mathcal{H}_i$, which are invariant under the action of LU or of SLOCC? Since local unitary transformations do not affect the structure of entanglement in the state $\ket{\psi}$, this question is important in quantum information.

    \item\label{multi_matrix} Counting multi-matrix gauge-invariant operators in matrix quantum mechanics \cite{deMelloKoch:2025ngs,deMelloKoch:2025rkw} or $\mathcal{N} = 4$ super Yang-Mills \cite{deMelloKoch:2020agz}. For instance, in $\mathcal{N} = 4$, local composite operators are scalars under $U(N)$ gauge transformations which are constructed from a string of matrix-valued fields, such as $F_{1, i_1}^{j_1} \ldots F_{n, i_n}^{j_n}$. In order to enumerate the number of independent gauge-invariant operators, one must therefore study the invariants that can be constructed from a collection of matrices. Such analyses often rely upon the result of \cite{PROCESI198763}, which proved that invariants of a collection of $N \times N$ matrices are generated by traces of products of individual matrices, and all relations between those traces follow from a generalization of the Cayley-Hamilton theorem. However, finding a full classification of the independent invariants in specific cases can be challenging.

\end{enumerate}

Although the problem of explicitly enumerating invariants is ubiquitous and important, it is very hard to solve in the general case. In this work, rather than approaching this problem from a rigorous mathematical perspective, we will take a more data-driven approach. Given a collection of objects transforming in some representation of a Lie group $G$, suppose that we can parameterize these objects with a collection of numerical labels. For instance, in the case of a single tensor transforming in a representation of the Lorentz group, one can simply choose these numerical labels to be the components of the tensor in any basis. We then generate many numerical instances of this tensor by randomly drawing components from some probability distribution, such as independent and identically distributed Gaussians. One can systematically enumerate the ways to form scalars from these numerical instances, which in general involves considering all possible ways of contracting the indices on the tensor with some set of invariant tensors associated with the group, such as the Kronecker delta $\tensor{\delta}{^\mu_\nu}$ in the case of a special orthogonal group action. Finally, one seeks relations between the different contractions by compiling the different numerical scalars across different random draws and performing numerical linear algebra. This step is carried out by listing all scalars at a fixed order -- that is, those invariants which involve the same number of instances of the numerical tensor -- and testing for linear combinations of these scalars which vanish. This process detects, for instance, Cayley-Hamilton-type formulas which relate traces of powers of a matrix to products of lower-order traces.

This procedure can be repeated until it ``stabilizes'' in the sense that one fails to find any new invariants for several consecutive orders. Of course, this does not represent a mathematical proof that the list of scalars generated by this procedure is functionally complete. Nonetheless, it offers an experimental approach that produces a conjectural set of independent invariants and can increase one's confidence that this set is complete. It also provides an explicit mechanism for identifying functional relations between invariants.

There are two primary goals of the present work:

\begin{enumerate}[label = (\Alph*)]

    \item We will describe the algorithm which was briefly sketched above in more complete detail. We explain how this algorithm can be used to experimentally study the independent invariants that can be constructed from any collection of indexed objects transforming in some representation of a Lie group, including (in principle) the three motivating examples \ref{stringy}, \ref{quantum}, and \ref{multi_matrix} that we have mentioned.

    \item\label{three_form_case_study} We apply this algorithm in the extended case study of the invariants that can be constructed from an antisymmetric $3$-form $H_{\mu \nu \rho}$ in six spacetime dimensions. This allows us to answer the natural analogue of the question posed above -- namely, what is the most general Lagrangian for an antisymmetric field strength $F_{\mu \nu}$ in four dimensions -- in the $6d$ setting. This is well-motivated from string theory since, for instance, the Kalb-Ramond $2$-form $B_{\mu \nu}$ with field strength $H_{\mu \nu \rho}$ is part of the field content of the NS-NS sector of string theory, and likewise an antisymmetric $3$-form field strength describes the gauge dynamics on the worldvolume of an M5-brane. We solve this problem in three ways, finding three different parameterizations which are analogous to the three choices (\ref{x1_and_x2}), (\ref{S_and_P}), (\ref{spinor_variables}) of independent scalars for a $2$-form in $4d$, and present relations between two of these sets of variables. The application of our algorithm in each of these three settings introduces different technical subtleties, but in all cases we find that there are five independent invariants, and thus the most general Lagrangian for a field $H_{\mu \nu \rho}$ in $6d$ is a function of five variables.
\end{enumerate}

Despite the fact that we focus on only the three versions of the case study \ref{three_form_case_study} in this work, we emphasize that the algorithm proposed here is quite general, and it is our hope that the present article will inspire a data-driven approach to studying invariants of other tensors of interest in mathematics and physics.

The structure of this paper is as follows. In Section \ref{sec:review}, we review some generalities about the theory of invariants, and contrast our approach with several other common techniques for studying invariants of tensors. Section \ref{sec:algorithm} presents the details of our proposed algorithm. In Section \ref{sec:threeform}, we carry out the application of our method to solving the problem of enumerating the independent invariants that can be constructed from a $3$-form in six dimensions in three different ways, which again parallel the three descriptions (\ref{x1_and_x2}), (\ref{S_and_P}), (\ref{spinor_variables}) of the scalars that can be built from a $2$-form field strength in $4d$. Section \ref{sec:conclusion} summarizes our results and presents directions for future research. We have relegated some data regarding graphs that represent contractions, and relations between invariants, to Appendix \ref{app:relations}.

\textbf{Note added}. While this work was being completed, the interesting paper \cite{Cederwall:2025ywy} appeared, which studies invariants of tensors from an analytical (rather than numerical) perspective.

\section{Review of Invariant Theory}\label{sec:review}

We begin by giving an overview of some ingredients in the subject of invariant theory, for the benefit of the unfamiliar reader. See also \cite{Olver_1999,goodman2000representations} for more detailed introductions.

\subsubsection*{\ul{\it General Theory of Invariants}}

It may be instructive to offer a motivating example which illustrates why one might expect this problem to be difficult in general. Suppose that we are given a tensor, with some number of indices and some symmetry structure, which transforms under $SO(n)$ rotations. We wish to count the number of independent pieces of rotation-invariant data contained in this tensor. A natural conjecture is that one can do this by first counting the number of independent components of the tensor, and then subtracting the number of rotation generators of $SO(n)$, since we are free to perform rotations in order to set some number of tensor components to zero. For instance, a real symmetric $n \times n$ matrix
\begin{align}
    M_{ij} = M_{ji}
\end{align}
has $\frac{1}{2} n ( n + 1 )$ independent entries. We know that any such matrix can be diagonalized by a rotation matrix, so the $n$  resulting diagonal entries (the eigenvalues) represent the desired pieces of rotation-independent data. And indeed, subtracting the count $\frac{1}{2} n ( n - 1 )$ of generators of $SO(n)$ from the number of independent components yields
\begin{align}
    \frac{1}{2} n ( n + 1 ) - \frac{1}{2} n ( n - 1 ) = n \, ,
\end{align}
which correctly matches the number of rotation invariants that can be constructed from a real symmetric $n \times n$ matrix.

Unfortunately, this na\"ive counting prescription can fail in other cases. For instance, now consider the case of a self-dual three-form $G_{\mu \nu \rho}$ in six spacetime dimensions with Lorentz signature. By self-dual, we mean that this tensor obeys
\begin{align}\label{self_duality}
    G_{\mu \nu \rho} = \frac{1}{6} \epsilon_{\mu \nu \rho \sigma \tau \kappa} G^{\sigma \tau \kappa} \, ,
\end{align}
where $\epsilon_{\mu \nu \rho \sigma \tau \kappa}$ is the Levi-Civita symbol, and indices are raised or lowered with the flat Minkowski metric $\eta_{\mu \nu}$. 

Repeating the above argument, we first note that a generic $3$-form in six dimensions has ${6 \choose 3} = 20$ independent components, and for a tensor obeying (\ref{self_duality}), half of these entries are determined in terms of the others, leaving $10$ independent components. However, the number of Lorentz generators in six dimensions is $\frac{6 \cdot 5}{2} = 15$, so a na\"ive subtraction gives
\begin{align}
    10 - 15 = - 5
\end{align}
independent Lorentz scalars, which is clearly absurd. In fact, it is known that there is a single Lorentz scalar that can be constructed from a self-dual three-form in six dimensions.\footnote{This fact appears to be well-known, but a particularly nice discussion may be found in Section 4.3 and Appendix B.1 of \cite{Avetisyan:2022zza}.}

As this example makes apparent, the problem of counting independent invariants is more subtle -- for a particular representation, such as the self-dual tensor in $6d$, there may be generators of the group action which generically lie in the stabilizer, and must therefore be excluded in a putative subtraction argument.

To address this subtlety, let us step back and formulate the problem of interest with somewhat more generality and abstraction. Consider a vector space $V$ over a field $K$, which in our cases will always be either the real or complex numbers, and suppose that $V$ carries a representation of a Lie group $G$. For instance, $V$ may be the vector space whose basis elements are abstract components of a tensor. Let $K [ V ]$ be the ring of polynomial functions on $V$. There is a natural action of $G$ on $K [ V ]$, induced by the representation of $G$ on $V$. At the risk of being pedantic, let us give an explicit example: if one thinks of $K [ V ]$ as the collection of polynomials in formal variables which represent components of a tensor $T_{i_1 \ldots i_k}$ in a basis, and $R \in G = SO ( n )$ simply rotates these components as 
\begin{align}
    T_{i_1 \ldots i_k} \to T_{i_1 \ldots i_k}' = \tensor{R}{^{j_1}_{i_1}} \ldots \tensor{R}{^{j_k}_{i_k}} T_{j_1 \ldots j_k} \, ,
\end{align}
then each monomial $\left( T_{i_1 \ldots i_k} \right)^n$ in tensor components transforms by taking the appropriate power $\left( T_{i_1 \ldots i_k}' \right)^n$ of the rotated components, and the group action extends to arbitrary polynomials by linearity and acting trivially on scalars in $K$.\footnote{Strictly speaking, $K [ V ]$ is typically defined in terms of polynomials in elements of the \emph{dual} space $V^\ast$ of $V$, but we will not carefully emphasize this distinction in what follows.} More generally, given $g \in G$ and $f \in K [ V ]$, let $f^g$ represent the result of performing this natural group action on the polynomial $f$. Then one can consider the subring of invariant polynomials,
\begin{align}\label{invariant_ring_definition}
    K [ V ]^G = \{ f \in K [ V ] \; \mid \; f^g = f \text{ for all } g \in G \} \, .
\end{align}
In words, each element of $K [ V ]^G$ represents a singlet which is constructed from elements of $V$ but which does not transform under the action of the group $G$. In the explicit example above, one such invariant element is associated with the contraction
\begin{align}
    T_{i_1 \ldots i_k} T^{i_1 \ldots i_k} \, ,
\end{align}
which is manifestly inert under rotations.

In this work, we will always suppose that the field $K$ is of characteristic zero -- again, like the real or complex numbers -- and we will make the technical assumption that $G$ is a linearly reductive group which acts rationally on $V$. These conditions apply to all situations of our interest, such as the rotation group and Lorentz group (more precisely, its complexification as an algebraic group). Under these assumptions, one can say more about the structure of the ring, or algebra,\footnote{Whether we regard the collection of invariants as a ring or an algebra only affects the role of the base field $K$, and as we will typically specialize to $K = \mathbb{R}$ or $K = \mathbb{C}$, the difference is immaterial. We will therefore use the terms interchangeably in what follows.} of invariants defined in (\ref{invariant_ring_definition}).

First, it is known that the algebra of invariants is finitely generated, which is Hilbert's finiteness theorem. That is, every invariant can be written as a polynomial in a finite number of generators with coefficients in $K$. This may not come as a surprise in the motivating example of invariants of tensors, since such tensors have finitely many components in any basis, so one might naturally expect that the number of functionally independent invariants should also be finite.

However, the fact that the invariant algebra (\ref{invariant_ring_definition}) can be built from a finite list of generators does not preclude the possibility that there may be functional relations between the invariants in the generating set. To constrain the possible form of such relations, it is convenient to discuss some further substructure in the invariant algebra (\ref{invariant_ring_definition}) which involves a decomposition into a set of \emph{primary invariants} $\theta_1 , \ldots , \theta_p$ and \emph{secondary invariants} $\eta_1 , \ldots, \eta_s$. Let us first discuss the primary invariants, which are sometimes also referred to as a \emph{homogeneous system of parameters} (HSOP). By Noether normalization, there always exist a set of polynomial invariants $\theta_i \in K [ V ]^G$ which are algebraically independent (i.e. there is no polynomial equation which relates the $\theta_i$ to one another) and such that the collection of $\theta_i$ simultaneously vanish on a locus known as the nullcone. One can therefore consider the ring of polynomial functions in all of these primary invariants,
\begin{align}
    P = K [ \theta_1, \ldots, \theta_p ] \, .
\end{align}
Although each element of $P$ is clearly an invariant, and thus lies in $K [ V ]^G$, in general $P$ will only be a subring of (\ref{invariant_ring_definition}), and therefore there will exist some invariants that cannot be expressed as polynomials in the $\theta_i$. At this stage, one can only conclude that the full ring or algebra of invariants $K [ V ]^G$ is a finitely-generated module over $P$.

Further conclusions may be drawn under the assumption that $K [ V ]^G$ is a \emph{Cohen-Macaulay ring}, which is another technical condition.\footnote{This notion is inspired by the work of Cohen \cite{Cohen1946CompleteLocalRings} and Macaulay \cite{Macaulay1916ModularSystems}, although neither introduced the modern definition; see \cite{bruns1998cohen} for a more recent introduction.} Fortunately, due to the Hochster-Roberts theorem \cite{HochsterRoberts1974}, it is known that $K [ V ]^G$ is always Cohen-Macaulay if the group $G$ is linearly reductive, which we have already assumed. Therefore, for the purposes of this work, every invariant ring obeys the Cohen-Macaulay condition, which in particular implies that it admits a Hironaka decomposition\footnote{It is stated in \cite{nagata1962local} that this decomposition was introduced in Hironaka's unpublished master's thesis in 1962, for which we were unable to find a reference, but see \cite{derksen2002computational} for a textbook discussion.} that takes the form
\begin{align}\label{hironaka}
    K [ V ]^G = \bigoplus_{k=1}^{s} \eta_k P \, ,
\end{align}
for a set of additional invariants $\eta_k$, which are the secondary invariants mentioned above.

Although the choice of primary invariants $\theta_i$ for a given invariant ring is not unique, the content of (\ref{hironaka}) is that -- once we have fixed a particular choice of primary invariants -- every other invariant $x \in K [ V ]^G$ can be expressed uniquely as
\begin{align}
    x = \sum_{k=1}^{s} \eta_k f^k \, , \qquad f^k \in P = K [ \theta_1, \ldots, \theta_p ] \, .
\end{align}
That is, every invariant is a unique linear combination of terms, each of which involves a secondary invariant multiplied by a polynomial in the primary invariants.

One might again ask about relations between invariants. Although we have already emphasized that the $\theta_i$ are algebraically independent, every product of secondary invariants satisfies a relation of the form
\begin{align}\label{secondary_relation}
    \eta_k \eta_m = \sum_{j = 1}^{s} \eta_j f^j_{k m} \, , \qquad f^j_{k m} \in P \, ,
\end{align}
which is reminiscent of the structure of a Lie algebra. 

To summarize what we have discussed thus far, the Hironaka decomposition corrects the na\"ive counting argument which motivated this review in two ways. First, the number of primary invariants -- sometimes known as the \emph{Krull dimension} -- need not agree with a straightforward enumeration which subtracts the number of symmetry generators from the number of degrees of freedom. Rather, the Krull dimension gives a corrected count
\begin{align}
    p = \dim ( V ) - \left( \dim ( \mathfrak{g} ) - \dim ( \mathfrak{h} ) \right) \, ,
\end{align}
where $\mathfrak{g}$ is the Lie algebra of $G$ so that $\dim ( \mathfrak{g} )$ is the number of generators, and $\mathfrak{h}$ represents the collection of generators that are in the typical stabilizer and thus act trivially on elements of $V$. For the example of the self-dual $3$-form mentioned above, it turns out that $\dim ( \mathfrak{h} ) = 6$ so the Krull dimension is $10 - ( 15 - 6 ) = 1$. Here there is a single primary invariant $\theta_1$ and only the trivial secondary invariant $\eta_1 = 1$, and all invariants are functions of $\theta_1$. Second, the na\"ive counting has been further corrected by the observation that, in general, not every element of $K[V]^G$ is simply a polynomial in the primary invariants, but rather secondary invariants may be needed in certain cases.\footnote{See \cite{Gripaios_2021,gripaios2020lorentzpermutationinvariantsparticles,Lester:2020jrg} for an application of the formalism reviewed in this section to studying Lorentz- and permutation-invariant combinations of particle momenta in scattering processes.}

\subsubsection*{\ul{\it Other Approaches to Studying Invariants}}

As we have emphasized, the philosophy of the present work is to take a data-driven approach to classifying the invariants of tensors. Having reviewed the relevant generalities on the theory of invariants, we will now discuss other approaches to enumerating invariants which have been used in the physics literature in order to contrast them with our method.

One set of analytical, rather than numerical, techniques for counting invariants involve certain generating functions that resemble partition functions and which are computed using the \emph{Molien-Weyl formula}. This method can be applied in both a ``fine-grained'' form which keeps track of both the numbers of primary and secondary invariants, or in a ``coarse-grained'' form which simply counts the total number of invariants at each order. In the coarse-grained case, it is common to refer to the resulting generating function
\begin{align}\label{hilbert_series}
    Z ( t ) = \sum_n b_n t^n
\end{align}
as a \emph{Hilbert series} or \emph{Hilbert-Poincar\'e series}. Here $b_n$ is the total number of independent invariants that can be constructed from contractions of $n$ copies of the tensor of interest.

Both the coarse-grained (see e.g. \cite{Cremonini:2022cdm}) and fine-grained (for instance, \cite{deMelloKoch:2025qeq}) versions of this generating function technique have been applied in the theoretical physics literature. Let us briefly review the fine-grained version. In this case, the corresponding generating function takes the form
\begin{align}
    Z ( t ) = \frac{\sum_k t^{\deg ( \eta_k ) } }{\prod_i \left( 1 - t^{\deg ( \theta_i  ) } \right) } \, ,
\end{align}
where the sum in the numerator runs over all secondary invariants $\eta_k$, and the product in the denominator runs over all primary invariants $\theta_i$. Here $\deg$ refers to the degree of a given invariant, e.g. how many copies of a tensor appear in the contraction which defines the corresponding invariant. If we instead define $p_i$ as the number of primary invariants at order $i$ and $s_k$ as the number of non-trivial secondary invariants at order $k \geq 1$, so that $\sum_i p_i = p$ and $1 + \sum_{k \geq 1} s_k = s$ in our earlier notation, and strip off the trivial secondary invariant $\eta_1 = 1$, this generating function can also be written as
\begin{align}\label{fine_grained}
    Z ( t ) = \frac{1 + \sum_{k \geq 1} s_k t^k}{\prod_i ( 1 - t^i )^{p_i}} \, .
\end{align}
We therefore see that the structure of (\ref{fine_grained}) encodes information about both the counts of primary and secondary invariants. If one had discarded the structure of the numerator and denominator in (\ref{fine_grained}), simply expanding this expression as a power series in $t$, one would recover the coarse-grained expression (\ref{hilbert_series}) which counts all invariants at a given order. Depending on whether one is interested in the precise decomposition into primaries and secondaries, or only on the general count of all invariants, one or the other of these forms for the generating function may be more convenient.

For Lie groups, the generating function $Z ( t )$ may be computed via an integral expression provided by the Molien-Weyl integral formula, which has been applied several times in the physics literature \cite{Pouliot:1998yv,Romelsberger:2005eg,Hanany:2007zz,Dolan:2007rq,Dolan:2008qi,Hanany:2008sb} (although sometimes not using this name) and takes the form
\begin{align}\label{molien_weyl}
    Z ( t ) = \int \frac{d \mu ( g ) }{\det \left( 1 - t R ( g ) \right)} \, ,
\end{align}
where $d \mu ( g )$ is the Haar measure on the group $G$, and $R$ is the representation under consideration. In certain cases, one can evaluate this integral in closed form, for instance by integrating over the maximal torus and including a Vandermonde determinant factor. This gives a mathematically rigorous way of establishing the number of invariants which occur at a given order, but it does not provide explicit expressions for these invariants, nor does it reveal how to express a new (non-independent) invariant in terms of the independent ones. In contrast, although we have said that our approach is numerical and thus non-rigorous, it \emph{does} provide such explicit expressions and relations.

Another familiar way to analyze invariants of tensors is to decompose products of representations into irreducible representations and look for singlets in this decomposition. This procedure may be carried out programmatically using software like LiE \cite{vanLeeuwenCohenLisser_LiE_1992} or LieART \cite{Feger:2012bs,Feger:2019tvk}. For instance, using the LieART package in Mathematica, one can enter the command

\medskip

\begin{mathin}
DecomposeProduct[Irrep[SO6][10], Irrep[SO6][10]]
\end{mathin}

\begin{mathout}
$\irrep[1]{20} + \irrep{35} + \irrep{45}$
\end{mathout}

\medskip

Here the function \texttt{Irrep[SO6][10]} identifies the $10$-dimensional irreducible representation of $\mathrm{SO} ( 6 )$, which is the self-dual tensor.\footnote{Strictly speaking, for a real self-dual tensor in Lorentz signature we should use $\mathrm{SO} ( 1, 5 )$, but here we will be cavalier about the distinction as it does not affect the results of the present discussion.} Since there is no singlet in this decomposition, we can conclude that there is no invariant that can be constructed from two copies of a self-dual $3$-form in six dimensions. However, taking four copies yields

\medskip \medskip

\begin{mathin}
DecomposeProduct[Irrep[SO6][10], Irrep[SO6][10], Irrep[SO6][10], Irrep[SO6][10]]
\end{mathin}

\medskip \medskip

\begin{mathout}
$\irrep{1} + 3 ( \irrep{15} ) + 2 ( \irrep[1]{20} ) + \irrep{35} + 3 ( \irrep{45} ) + 3 ( \irrepbar{45} ) + 6 ( \irrep{84} ) + 3 ( \irrep{105} ) + \irrep{165} + 7 ( \irrep{175} ) + 3 ( \irrep{189} ) + 8 ( \irrep{256} ) + 6 ( \irrep{280} ) + 3 ( \irrep{315} ) + 6 ( \irrep[1]{360} ) $
\end{mathout}

\medskip \medskip

In this case, we do see a singlet $\irrep{1}$ in the tensor product decomposition, which correctly identifies that there is a single quartic invariant that can be constructed from a self-dual tensor in six dimensions. The irrep decomposition is useful for counting invariants that can occur at a fixed order, but like the Molien-Weyl approach, it differs from our technique in several ways. This decomposition is again an exact/theoretical method rather than a numerical one; it only identifies the existence of invariants at a given order, but says nothing about polynomial dependence or independence; and this strategy does not give explicit expressions for the index contractions that form the given invariants, nor equations which relate dependent invariants to those in an independent generating set.

Finally, let us mention a few other approaches to this problem. Much as the eigenvalues of a matrix form good examples of basis-independent data, there is a sizable literature on generalizations to eigenvalues of tensors; see, for instance, the textbook \cite{Qi2018TensorEA} and references therein, or the sampling of works \cite{QI20051302,abo2015eigenconfigurations,2010arXiv1004.4953C}. Although tensor eigenvalue techniques have been applied in theoretical physics \cite{Sasakura:2024awt,Delporte:2024izt}, there are multiple definitions of such eigenvalues -- some of which apply only to tensors with additional symmetries -- and it is fair to say that the theory is less well-understood than the corresponding techniques for matrices. Although it would be interesting to relate the numerically identified invariants constructed here to tensor eigenvalues, we will not attempt this in the current article.

One step of our algorithm will involve the construction of certain labeled graphs that represent tensor contractions. Similar graphical approaches have also been used in many other works; see, for instance, Appendix B of \cite{Chandra:2023afu} for invariants constructed from $3$-tensors representing intersection numbers on Calabi-Yau threefolds, or papers like \cite{Avohou:2019qrl,Avohou:2024agh} which also count invariants by enumerating particular graphs. Although we also employ graphical representations, the distinction is that our approach also generates many randomly-drawn instances of numerical tensors (and hence the invariants represented by such graphs) with the goal of discovering relations between invariants and thus an independent generating set.

\section{Description of Algorithm}\label{sec:algorithm}

We now turn to the primary method of interest in the present article. We refer to this approach both as ``data-driven'' and as a ``machine learning (ML)'' technique, where we use the term ML in a more general sense that includes any statistical approach which extracts structured information from data (rather than a more restrictive definition that might include only modern artificial intelligence architectures such as neural networks, transformers, etc.). Our view is that a more inclusive definition of ML should include, at minimum, various unsupervised learning techniques appearing in machine learning libraries like scikit-learn \cite{scikit-learn,sklearn_api}, such as principal component analysis (PCA), whose goal is to reduce the dimensionality of a dataset using linear algebra. Our algorithm is, in some sense, an invariant-theoretic version of PCA, insofar as it attempts to reduce a large collection of polynomials in tensor components to a smaller subspace representing distinguished elements of $K [ V ]^G$, which are invariant under the action of a Lie group $G$.

Fix a vector space $V$ transforming in a representation $R$ of a Lie group $G$. Typically we can regard elements of $V$ as objects which carry some number of indices. For instance, suppose that $G$ is linearly reductive (as we always assume) and $\rho$ is a faithful finite-dimensional representation. Then every finite-dimensional rational representation $R$ of $G$ is isomorphic to a subrepresentation -- hence, since $G$ is linearly reductive, a direct summand -- of a finite direct sum of representations of the form
\begin{align}
    \rho^{\otimes n} \otimes \left( \rho^\ast \right)^{\otimes m} \, ,
\end{align}
whose elements are tensors with $n$ upper and $m$ lower indices of $\rho$. A familiar example is the Riemann tensor $\tensor{R}{^\mu_\nu_\rho_\sigma}$, which lives in a subrepresentation involving one upper and three lower indices, each transforming in the vector or covector representation, and enjoying the usual symmetries of the Riemann tensor. More generally, a tensor may also involve spinor indices rather than vector indices.

We will first describe how our numerical approach is implemented in general, and then describe simplifications that occur in particular cases. Let $T$ be such an indexed presentation for the representation of interest, where the indices carried by $T$ have been suppressed. Enumerate any other invariant tensors $I_1, \ldots, I_m$ for the given representation. We will use the term \emph{contraction graph} or \emph{tensor network} for a vertex-labeled and edge-labeled multigraph, where each vertex is labeled as either $T$ or one of the invariant tensors $I_k$, $k = 1 , \ldots, m$, and each edge carries two integer labels indicating which of the indices on the two nodes that it connects are to be contracted. In cases where we distinguish between different types of indices (e.g. upper versus lower, or spinor versus vector), only a subset of labeled edges are permitted, corresponding to allowed contractions. For a fixed order $N$, we construct all possible contraction graphs with the following properties.

\begin{enumerate}[label = (\Roman*)]
    \item There are exactly $N$ instances of nodes which are labeled $T$ and any number of nodes which represent invariant tensors $I_k$.

    \item For every node in the graph, there is precisely one labeled edge emanating from the node for each of the indices on that node.

    \item\label{connected} The graph has only one connected component.
\end{enumerate}

Put more simply, we essentially begin by constructing all tensor networks involving $N$ tensors $T$, and some number of invariant tensors, for which all indices are contracted. The connectedness criterion \ref{connected} serves two purposes: it excludes trivial ``invariant tensor bubbles'' which would correspond to multiplication by constants like $\epsilon_{ijk} \epsilon^{ijk}$ which are independent of elements of $T$, and it rules out graphs which represent contractions that factorize into products of lower-order contractions (such as the square of the trace of a matrix $\left( \tensor{M}{^i_i} \right)^2$), which we already know are dependent upon lower-order invariants.

We represent all of our graphs using the NetworkX library in Python \cite{hagberg2008exploring}. Operationally, constructing all graphs with these properties is the most computationally intensive step of our procedure. The precise algorithm that one should use for this enumeration depends upon the specific nature of the tensor under consideration, but we will give more details about the graph construction schemes in the specific cases that we discuss below.

Suppose that the collection of all such contraction graphs at order $N$ has been constructed, and that the number of them is $G_N$. We must then test for linear dependence between the scalars represented by each of these tensor networks. To do this, we randomly generate numerical instances of the tensor $T$ by drawing components using random number generation routines in the numpy library \cite{harris2020array}. The exact probability distribution which one uses for the tensor components is not especially important; for instance, one could draw from independent and identically distributed Gaussians, but in practice we will instead draw all components independently from a uniform distribution. We generate $G_N$ such random numerical instances of the tensor $T$, and for each of these instances, we compute the $G_N$ scalars represented by each of the contraction graphs at this order using the opt\_einsum library \cite{Smith2018}. We assemble this data into a $G_N \times G_N$ matrix whose columns correspond to the contractions represented by each of the graphs, and whose rows correspond to different numerical draws. Finally, we compute the numerical rank and nullspace of the matrix via a singular value decomposition (SVD) and look for singular values below a tolerance. Any null vector in this nullspace represents a linear combination of columns which vanishes, i.e. a linear relation between the scalars represented by the $G_N$ graphs. If any such linear relations exist, we remove a subset of the dependent scalars and retain a maximal linearly independent subset of the $G_N$ contraction graphs.\footnote{In practice, we repeat this experiment several times to ensure that the same null vectors are found for each run of the test, but finding ``false positive'' null vectors or ``false negative'' omission of relations due to numerical coincidences is very unlikely when tensor components are drawn randomly.}

Beginning from $N = 1$, we repeat the above procedure for $N = 2$, $N = 3$, and so on. However, although we have removed putative invariants which satisfy polynomial relations with other new invariants at a fixed order, we have not yet identified relations between higher-order invariants and products of lower-order invariants. At this stage, one can proceed with our algorithm in one of two ways, depending on whether it is known that there are only primary invariants and the trivial secondary invariant $\eta_1 = 1$, or if we wish to test for the presence of both primary and non-trivial secondary invariants.

\makeatletter
\begin{enumerate}
    \item[(P)]\def\@currentlabel{P}\label{primary_algorithm} Suppose it is known that there are only primary invariants $\theta_1 , \ldots , \theta_p$. Then, as the procedure is carried out for increasing values of $N \geq 2$, potential invariants are identified at each order $N$, and before seeking invariants of order $N+1$ we first remove potential invariants of order $N$ via a second test for polynomial dependence on the lower-order invariants. To do this, we enumerate all products of the independent lower-order invariants that have been identified thus far, such that the total degree of the product is $N$. We again perform random numerical draws of the tensor under consideration, with a total number of draws equal to the sum of the number of new independent order-$N$ invariants, plus the number of appropriate products of lower-order invariants. We assemble these numerical values into a matrix and test its numerical rank. Any element of the nullspace represents a polynomial relation between the new order-$N$ invariants and products of lower-order invariants, so we remove a subset of the order-$N$ invariants until we arrive at a minimal set which are both independent of one another and of products of lower invariants. Note that this procedure may not yet identify all possible relations because there could be additional constraints at higher orders, which will be addressed later.

    \item[(S)]\def\@currentlabel{S}\label{secondary_algorithm} Alternatively, assume that we wish to identify both primary invariants and potentially non-trivial secondary invariants. In this case, we must iteratively test subsets of the currently-identified invariants for relations of the form (\ref{secondary_relation}), again by numerically drawing values of the tensors, constructing the numerical invariants, and seeking null vectors which represent a relation of the form $\eta_k \eta_m - \sum_j \eta_j f^j_{km} = 0$. If, for a given subset of invariants, we numerically discover such relations for all $k$ and $m$, then we identify this subset of invariants as secondary. We then repeat the algorithm described in (\ref{primary_algorithm}) to eliminate any new candidate primary invariants which satisfy polynomial relations involving lower primary invariants.
\end{enumerate}
\makeatother

The algorithm (\ref{secondary_algorithm}) is more computationally intensive as it requires successively checking subsets of invariants to test whether they are secondary, but we stress that in principle the numerical approach can be used in either case (\ref{primary_algorithm}) or case (\ref{secondary_algorithm}).

The above steps describe how our algorithm identifies new invariants, and then prunes dependent ones, at each order $N$. In practice, we continue performing this procedure for larger and larger $N$ until no new invariants are identified -- that is, all of the putative new invariants satisfy relations involving products of lower invariants -- for multiple consecutive orders. The algorithm then terminates at some $N_{\text{final}}$. As a last step, after termination and the production of a final set of conjectural invariants, we may perform an additional check which forms all products of the conjectural invariants at orders $N > N_{\text{final}}$, and again generate more numerical draws in order to seek higher-order polynomial relations between the invariants in our set. The purpose of this additional step is to increase our confidence that the final set of invariants returned by the algorithm is truly independent, and does not satisfy higher-order constraints. In this additional higher-order pruning step, one again tests for all polynomial relations in the case (\ref{primary_algorithm}) where there are only primary invariants, while one only tests for relations not of the form (\ref{secondary_relation}) if secondary invariants may exist.

So far, this algorithm is completely general and applies to any tensor with any number of indices. The generalization to multiple tensors $T_i$ is clear: in this case, one simply considers labeled graphs with different labels for each of the distinct $T_i$, and then iterates over the total order $N$ which counts the combined number of occurrences of all of the tensors.

Finally, let us comment that this approach can often be simplified to exploit additional structure which is present in special cases. For instance, suppose that we are interested in invariants constructed from a totally antisymmetric tensor, and the only invariant tensor which we consider is the Kronecker delta. In this case, we can dispense with the invariant tensor nodes and simply use \emph{weighted graphs} (rather than labeled graphs) involving only tensor nodes $T$, and where the weight of an edge represents the number of common indices that are contracted between any pair of tensors. This is sufficient since a contraction on any particular index of an antisymmetric tensor yields the same result as a contraction on any other index, up to a sign, which does not affect polynomial dependence. Likewise, in this setting, we can forbid self-edges from a tensor $T$ to itself, since contracting any pair of indices on an antisymmetric tensor gives zero. Similarly, if we are constructing invariants associated with a totally symmetric tensor and again only using the Kronecker delta, we may also use weighted graphs between only tensor nodes which represent the number of contracted indices, although in the symmetric case self-edges are allowed and non-trivial.

\subsubsection*{\ul{\it Examples}}

To illustrate the technique above, let us first explain how it is applied in some trivial examples where the structure of the invariants is completely well-understood. First consider a symmetric $2 \times 2$ matrix $M_{ij} = M_{ji}$. Indices will be raised or lowered with the Kronecker delta, so we do not carefully distinguish between upper and lower indices, although the natural index placement for a map from vectors to vectors is $\tensor{M}{^i_j}$. At order $N = 1$, there is only a single contraction graph that one can draw, which corresponds to the trace of $M$:
\begin{center}
    \includegraphics[width=0.55\linewidth]{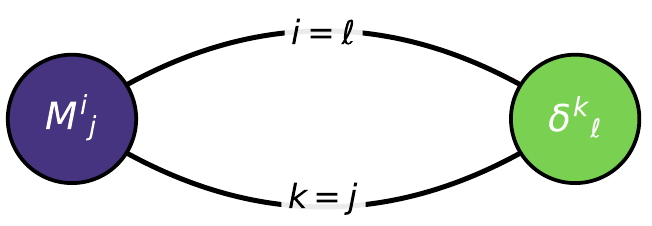}
\end{center}
For simplicity, we could also draw this contraction graph with the single node $\tensor{M}{^i_j}$ and a single self-edge labeled $i = j$ to represent the trace.

At order $N = 2$, there are two graphs that can be drawn, assuming that we may use both the invariant tensors $\tensor{\delta}{^i_j}$ and $\epsilon_{ij}$. The first is the trace of the square of the matrix,
\begin{center}
    \includegraphics[width=0.55\linewidth]{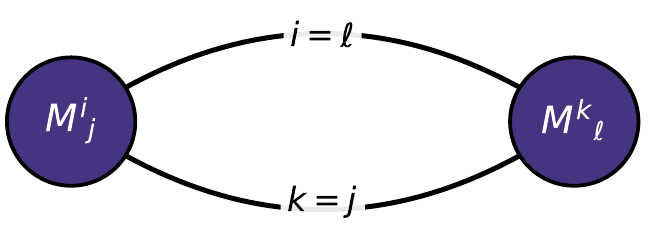}
\end{center}
In principle, we could also consider the trace of the product of $M$ and its transpose, but as $M$ is symmetric this does not give an independent invariant. 

The other invariant is built using the epsilon tensor and is twice the determinant of $M$. For simplicity, we will begin drawing abbreviated contraction maps where the contracted indices are already set equal on distinct nodes, and the corresponding edge simply labels the contracted index, as in the following: 
\begin{center}
    \includegraphics[width=0.65\linewidth]{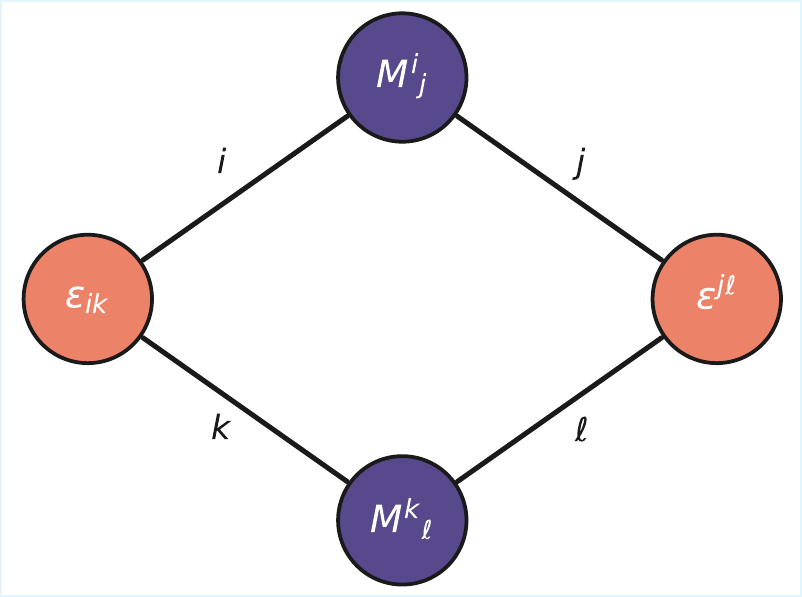}
\end{center}
Let the values of these three diagrams be
\begin{align}
    x^{(1)} = \tensor{M}{^i_i} \, , \qquad x_1^{(2)} = \tensor{M}{^i_j} \tensor{M}{^j_i} \, , \qquad x_2^{(2)} = \epsilon_{ik} \tensor{M}{^i_j} \epsilon^{j \ell} \tensor{M}{^k_\ell} \, ,
\end{align}
where the superscript indicates the order in $M$. We first draw numerical values for the $3$ independent entries of $M$ from a uniform distribution, compute the two invariants $x_1^{(2)}$ and $x_2^{(2)}$, and then repeat the numerical draw and computation a second time. This data is assembled into a $2 \times 2$ matrix, which we find to be full-rank, and therefore we find no linear relation between $x_1^{(2)}$ and $x_2^{(2)}$. This is expected, as the trace of the square of a $2 \times 2$ symmetric matrix is independent of its determinant. However, when we perform the second step of the algorithm and seek relations involving lower-order invariants, we find a null vector corresponding to the relation
\begin{align}
    \left( x^{(1)} \right)^2 - x_1^{(2)} - x_2^{(2)} = 0 \, ,
\end{align}
which is simply a consequence of the Cayley-Hamilton theorem for $2 \times 2$ matrices,
\begin{align}\label{cayley_hamilton_2d}
    M^2 - \left( \tr ( M ) \right) M + \det ( M ) \mathbb{I}_2 = 0 \, .
\end{align}
The numerical algorithm therefore correctly re-discovers the relations between invariants arising from (\ref{cayley_hamilton_2d}). Continuing to the final step of our algorithm, we proceed to $N = 3$ and above, in each case enumerating all of the tensor networks that can be constructed, drawing random numerical matrices, and testing for polynomial relations between the new invariants and appropriate powers of lower-order invariants. For every higher value of $N$ that we tested, all of the new invariants satisfy polynomial relations which determine them in terms of a minimal independent set of lower invariants, say $x^{(1)}$ and $x_1^{(2)}$. Again, this correctly reproduces the fact that every invariant constructed from a $2 \times 2$ matrix is a function of two variables, say the trace of the matrix and the trace of its square.

Other simple facts which we have reproduced using this algorithm include the following.
\begin{enumerate}[label = (\alph*)]
    \item\label{vector} There is only a single invariant associated with a vector $v_i$, namely its length-squared $v_i v^i$, and all other invariants are functions of this scalar.

    \item There are three invariants that can be built from a pair of vectors $v_i$, $w_i$, namely the squared lengths $v_i v^i$ and $w_i w^i$ along with the dot product $v_i w^i$.

    \item For an antisymmetric $(d-1)$ form in $d$ dimensions, there is also a single independent invariant, namely the quadratic combination with all indices contracted. This is clear since a $(d-1)$-form is Hodge dual to a vector, so this result must match case \ref{vector}.

    \item\label{antisymmetric_matrix} For an antisymmetric matrix in $d$ dimensions, there are $\lfloor \frac{d}{2} \rfloor$ independent invariants, where $\lfloor \, \cdot \, \rfloor$ is the floor function.

    \item Likewise, for a $(d-2)$-form in $d$ dimensions, there are also $\lfloor \frac{d}{2} \rfloor$ independent invariants, as such a form is Hodge dual to an antisymmetric $2$-form and thus coincides with \ref{antisymmetric_matrix}.
\end{enumerate}
Although all of these results are obvious, they serve as a check that our numerical implementation is correct. We now direct our attention to a slightly less trivial example.

\section{Three-Form in Six Dimensions}\label{sec:threeform}

We will now carry out the main application of our algorithm which is of interest in the present work, namely the enumeration of independent invariants that can be constructed from a generic $3$-form $H_{\mu \nu \rho}$ in six dimensions. It was recently shown, using generating function techniques, that there are five such independent invariants \cite{Cederwall:2025ywy}. In this case, the na\"ive counting prescription which we introduced in Section \ref{sec:review} gives the correct answer: there are ${6 \choose 3} = 20$ components associated with a generic $3$-form in $d = 6$ and $\frac{6 \cdot 5}{2} = 15$ Lorentz or rotation generators, so the subtraction $20 - 15 = 5$ yields the right number of invariants, and the generic stabilizer $\mathfrak{h}$ is trivial. It is also known that, for this problem, there is only a single secondary invariant $\eta_1 = 1$, so we may proceed with the simplest instantiation (\ref{primary_algorithm}) of our algorithm which only seeks primary invariants.

Although the number of invariants for this problem is known, to the best of our knowledge, no explicit formulas for the index contractions that yield a homogeneous system of parameters (i.e. collection of primary invariants) have yet been presented in the literature. As we reviewed in Section \ref{sec:intro}, having access to such explicit expressions can be useful for studying the properties of generic Lagrangians, as the parameterizations (\ref{x1_and_x2}), (\ref{S_and_P}), (\ref{spinor_variables}) are helpful in studying theories of non-linear electrodynamics. Furthermore, we are not aware of any reference that gives formulas relating other invariants to those in a given HSOP, nor relating elements in one convenient HSOP to those in another. We therefore find it worthwhile to apply our algorithm to this case, both to further illustrate the generality of our approach, and to provide some explicit formulas of the type we have just discussed.

\subsection{Trace Variables}\label{sec:trace}

The number of independent invariants at various orders depends on which invariant tensors one is permitted to use for contraction of indices. For instance, we saw in the case of a $2$-form in $4$ dimensions that the two invariants occur at orders $2$ and $4$ in the trace variables (\ref{x1_and_x2}), but both invariants are order $2$ in the Hodge dual variables (\ref{S_and_P}). In this section, we will construct the analogue of the trace variables (\ref{x1_and_x2}) for a three-form in six dimensions, finding five independent invariants at orders $2, 4, 4, 6, 8$.\footnote{If we may use all invariant tensors, the invariants occur at orders $2, 4, 4, 4, 6$, as pointed out in \cite{Cederwall:2025ywy}. This is related to whether one is interested in invariance under $O(6)$ or only $SO(6)$.} 

We work in Euclidean signature, writing the components of the $3$-form as $H_{abc}$, and allow contractions only with the Kronecker delta $\tensor{\delta}{^a_b}$. As described above, this allows us to simplify the graph construction algorithm: we may use weighted graphs where edge weights represent the number of contractions between a pair of tensors. As no self-edges are permitted, at order $N = 1$, there is no valid contraction graph that one can draw.

At order $N = 2$, there is a single permissible graph,
\begin{center}
    \includegraphics[width=0.55\linewidth]{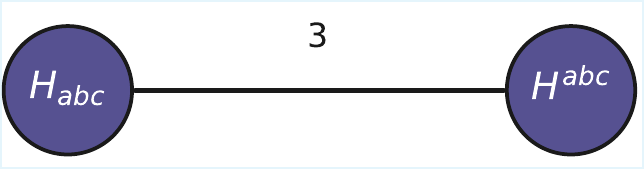}
\end{center}
which represents the contraction
\begin{align}
    x^{(2)} = H_{abc} H^{abc} \, .
\end{align}
To go to higher orders, it is necessary to implement a graph-theoretic algorithm which generates all valid contraction graphs. We do this using an exact backtracking enumeration of all weighted, undirected graphs with the property that each node has total degree equal to three. More specifically, the algorithm proceeds vertex-by-vertex, first numbering each vertex of the graph with an integer $v \in \mathbb{N}$. At each step, the algorithm computes the number of additional edges needed by a given vertex $v$, then enumerates all assignments of non-negative integers that represent feasible assignments of edges from $v$ to forward nodes $j > v$ without exceeding their edge budgets. Before recursing to later nodes, the algorithm first checks whether the total remaining capacity downstream can supply the necessary edges, and it prunes the current assignment if this check fails. For each feasible assignment, the algorithm performs in-place updates of the current edge assignments, then recurses to node $v + 1$, and so on. After finding a feasible assignment of all edge weights, the algorithm then backtracks and repeats the procedure. As this algorithm exhaustively checks all possible edge weight assignments, it is guaranteed to generate every valid contraction graph for a given number of tensor nodes. However, it will generically ``overcount'' by producing some tensor networks which are related by graph isomorphisms. To correct for this, in the final step our algorithm deduplicates such isomorphic graphs by converting each tensor network to an igraph \cite{ig1,ig2,ig3} object, testing for isomorphism using igraph's VF2 isomorphism routine, and retaining a maximal collection of non-isomorphic graphs.

Performing the graph algorithm described above at order $N = 4$, we find two inequivalent contractions represented by the tensor networks
\begin{center}
    \includegraphics[width=0.8\linewidth]{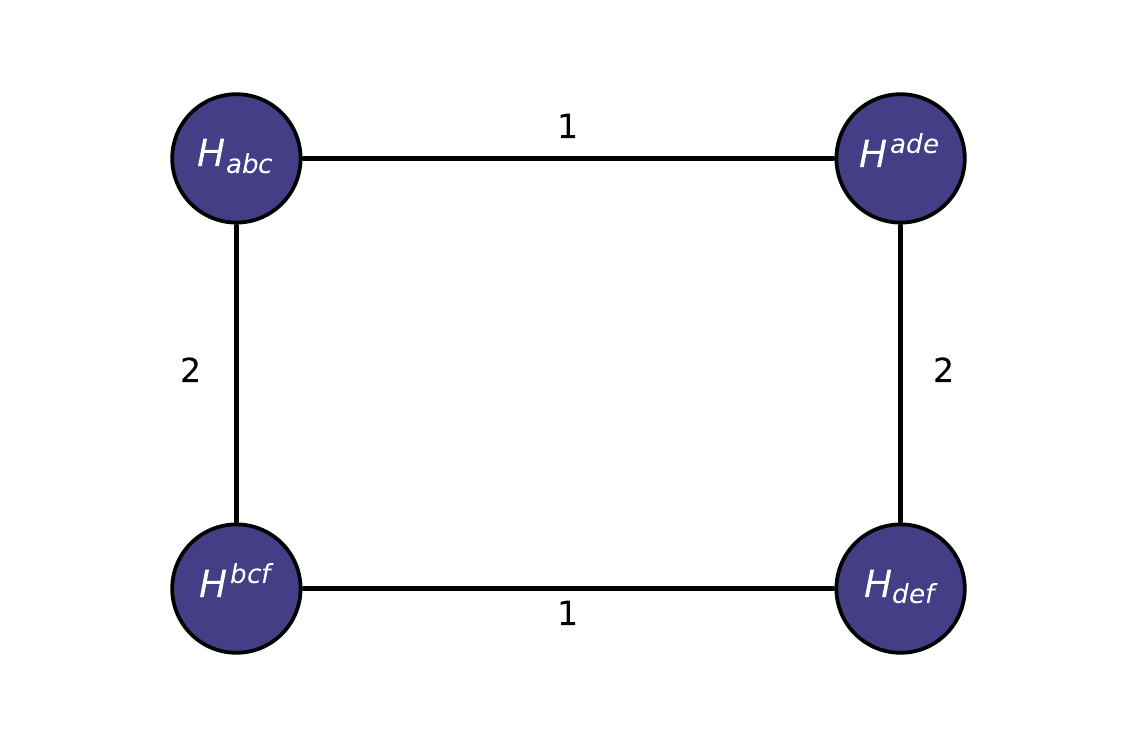}
\end{center}
and
\begin{center}
    \includegraphics[width=0.8\linewidth]{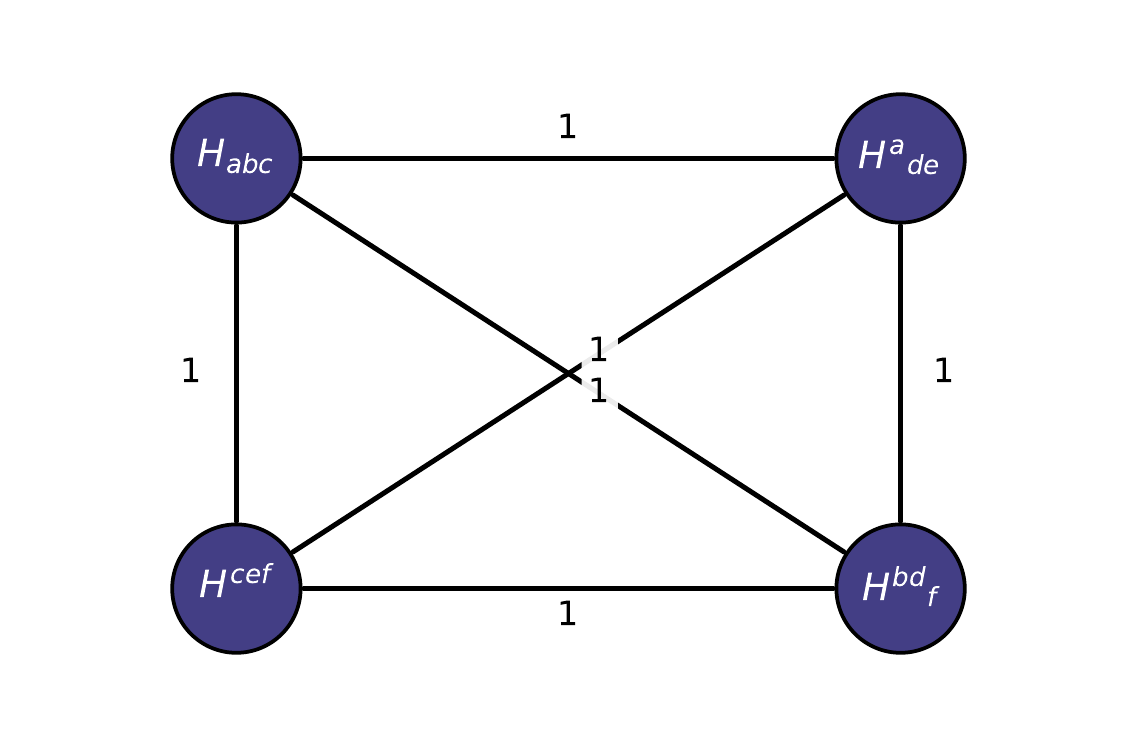}
\end{center}
Let the values of these contractions be
\begin{align}
    x^{(4)}_1 = H_{abc} H^{ade} H_{def} H^{bcf} \, , \qquad x^{(4)}_2 = H_{abc} \tensor{H}{^a_d_e} H^{cef} \tensor{H}{^b^d_f} \, .
\end{align}
We follow the steps of the algorithm of Section \ref{sec:algorithm}, testing first for linear relations between $x_1^{(4)}$ and $x_2^{(4)}$, and then testing for a relation between these two invariants and the square $\left( x^{(2)} \right)^2$ of the second-order invariant. In both cases, we find that the corresponding numerical matrix obtained by performing random draws of tensor components is full-rank, so there are no null vectors and therefore no relation between the invariants at this order.

Next we go to order $N = 6$. Our graph-theoretic algorithm outputs six inequivalent contraction graphs at this order. For the remainder of this section, we will not present every graph and polynomial relation explicitly, but one can find a sampling of additional graphs and relations in Appendix \ref{app:relations}. We perform the first step of checking for linear dependence between the six contractions at order $N = 6$ -- again by performing six random draws of the numerical tensor $H_{abc}$, computing the six contractions for each draw, and testing for null vectors of the resulting $6 \times 6$ matrix of numerical contractions -- and find that only three of these contractions are linearly independent. The three independent graphs are
\begin{center}
    \includegraphics[width=0.8\linewidth]{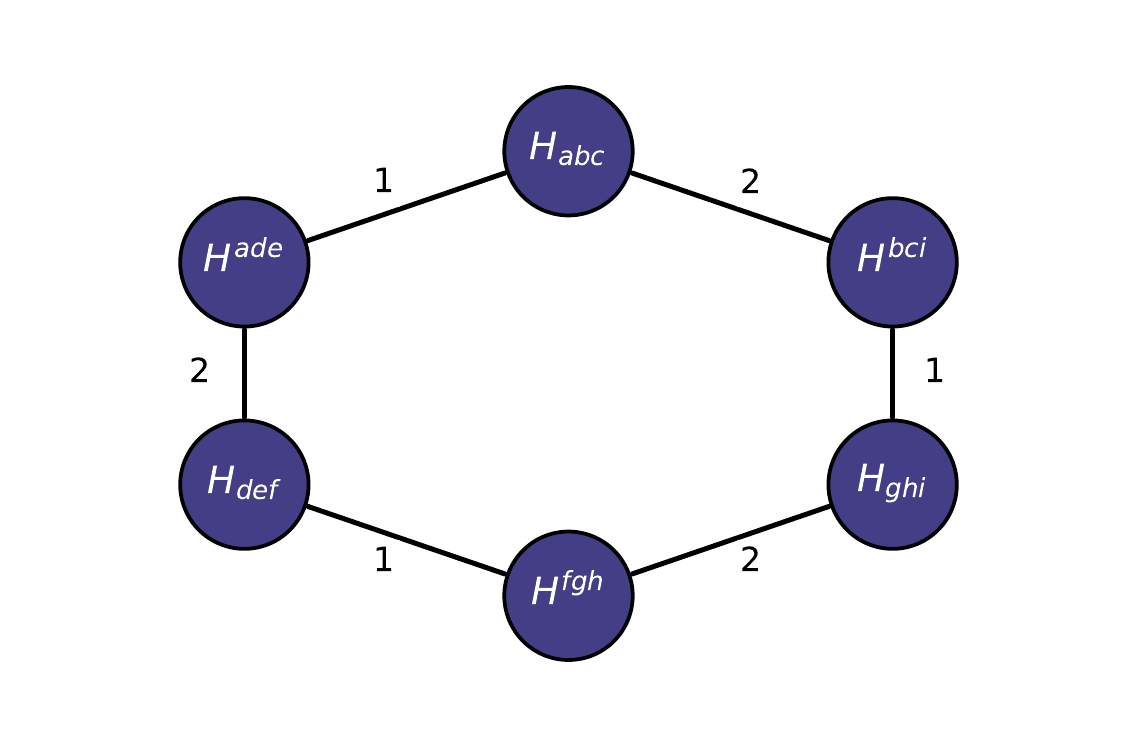}
\end{center}
and
\begin{center}
    \includegraphics[width=0.8\linewidth]{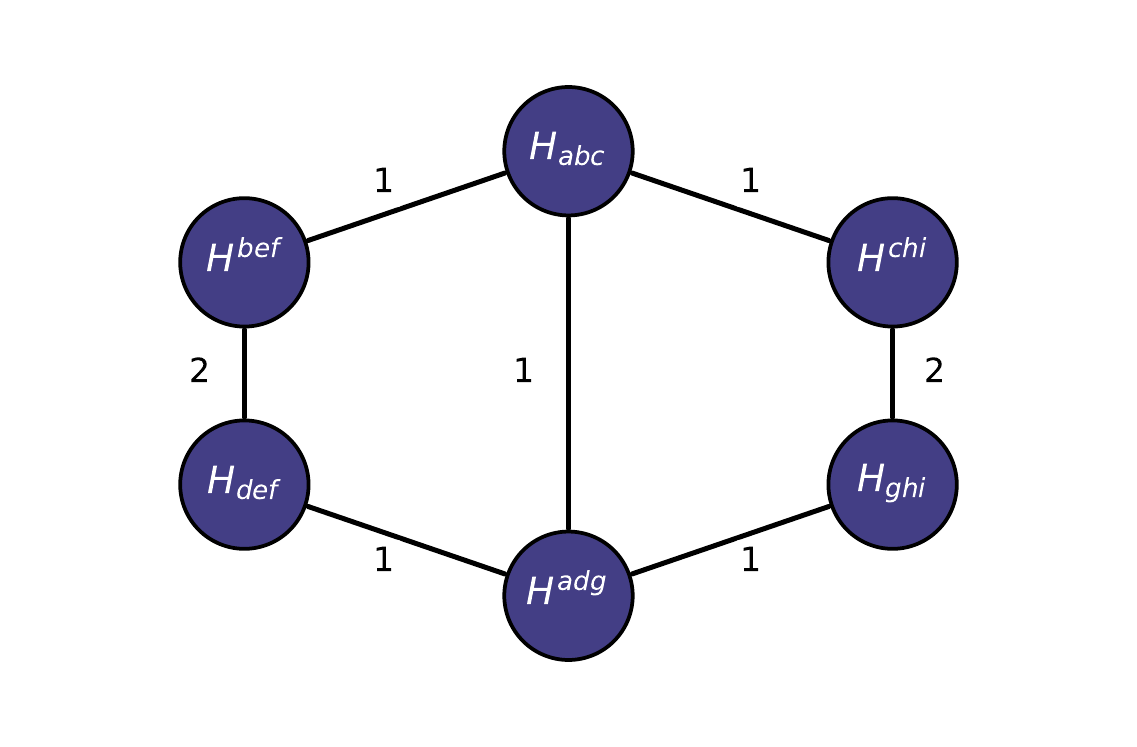}
\end{center}
and
\begin{center}
    \includegraphics[width=0.8\linewidth]{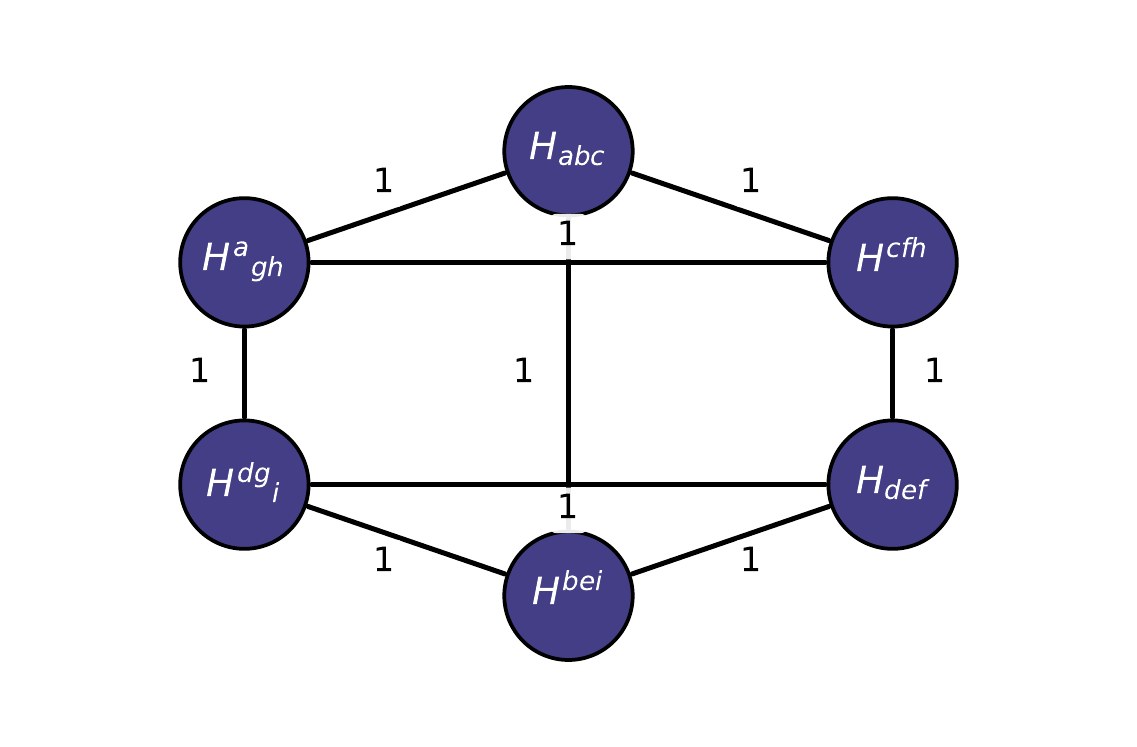}
\end{center}
However, when we perform the second step at this order, we find non-trivial relations between these three graphs and products of lower-order contractions. In fact, only one genuinely new invariant is introduced at this order, which we choose to be the second of the three graphs displayed above. Let this new invariant be
\begin{align}
    x^{(6)} = H_{abc} H^{chi} H_{ghi} H^{adg} H_{def} H^{bef} \, .
\end{align}
Next we go to order $N = 8$, where $20$ non-isomorphic contraction graphs are found. After testing for linear dependence, we find that only $6$ of these $20$ invariants are linearly independent of the others. We then test for polynomial relations among these $6$ invariants and products of lower-order invariants, and find that only one independent invariant is actually introduced at order $8$, which we choose to be the contraction graph
\begin{center}
    \includegraphics[width=0.8\linewidth]{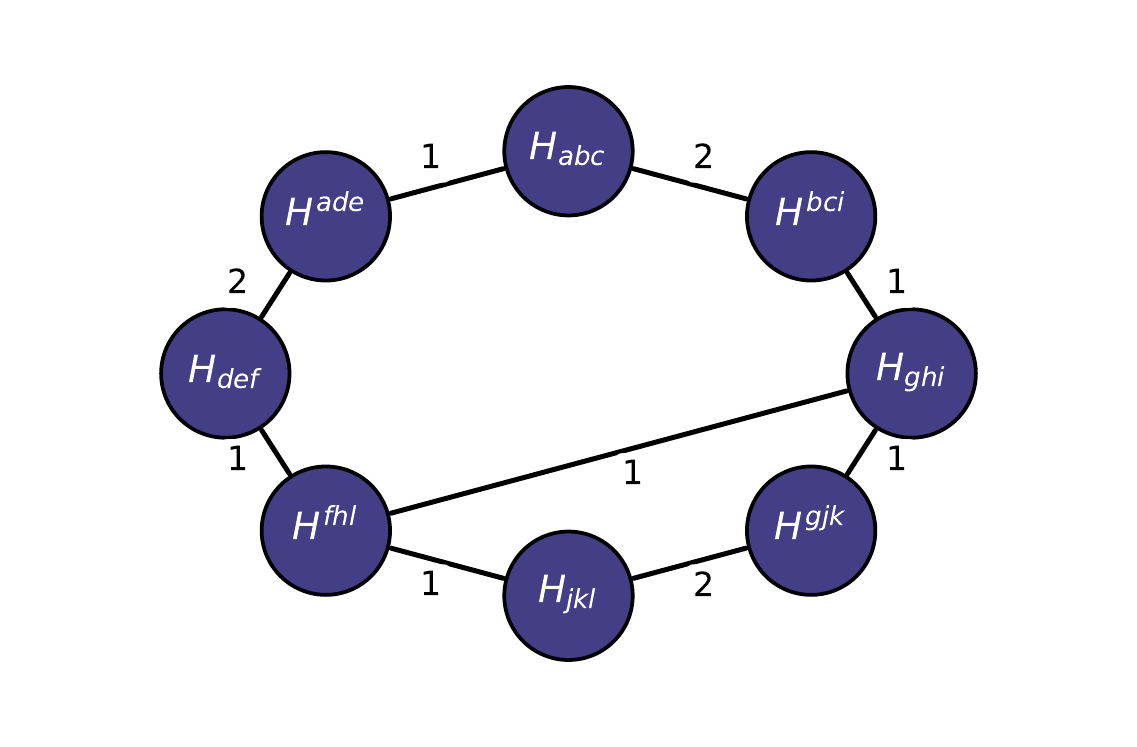}
\end{center}
associated with the invariant
\begin{align}
    x^{(8)} = H_{abc} H^{bci} H_{ghi} H^{gjk} H_{jkl} H^{fhl} H_{def} H^{ade} \, .
\end{align}
At this stage, we find that the algorithm terminates -- at any higher order in $N$, such as $N = 10$, all of the newly-generated invariants satisfy polynomial relations among themselves and products of lower-order invariants. This is to be expected from the observation of \cite{Cederwall:2025ywy} that there are only five independent invariants that can be constructed from a $3$-form in six dimensions, but now we have an explicit generating set:
\begin{align}\label{trace_variables_final}
    \left\{ x^{(2)} , x^{(4)}_1 , x^{(4)}_2 , x^{(6)} , x^{(8)} \right\} \, ,
\end{align}
whose expressions are given above.

The final step of our algorithm is to increase our confidence that these five invariants are truly independent, and do not satisfy any relations involving higher-order polynomials. We have checked up to order $N = 18$ that products of these invariants with total degree $N$ do not satisfy any such relation. This constitutes strong numerical evidence that the five variables (\ref{trace_variables_final}) are a complete set of independent invariants, and thus that the most general Lagrangian for a $3$-form $H_{abc}$ in six dimensional Euclidean spacetime is a function
\begin{align}
    \mathcal{L} = \mathcal{L} \left( x^{(2)} , x^{(4)}_1 , x^{(4)}_2 , x^{(6)} , x^{(8)} \right) \, .
\end{align}
Finally, let us make a numerological comment. In the procedure described above, we have retained only connected graphs, to avoid constructing redundant invariants which factorize into products of lower-order invariants. If we had instead constructed all allowed contraction graphs, we would have found (for instance) three tensor networks at order $N = 4$, including the two displayed above, as well as a third graph which is simply two disconnected copies of the unique $N = 2$ graph. We have also repeated the graph construction algorithm above in the case where one allows disconnected graphs and counted the number of distinct isomorphism classes of graphs at each even order in $N$, finding the counts $1, 3, 9, 32, 135, 709$ for $N = 2, 4, 6, 8, 10, 12$, respectively. This matches the first several terms in the sequence \cite{graphSequence} of the number of isomorphism classes of 3-regular loopless multigraphs of order $N = 2n$, providing a further check that our graph enumeration algorithm is correct. See \cite{travis1999graphical} for an analytical method for enumerating such graphs and generating this sequence.

\subsection{Hodge Dual Variables}\label{sec:hodge}

As a second application, we will now find another HSOP for the ring of scalars that can be constructed from a $3$-form in six Euclidean dimensions. Our construction will parallel the definition (\ref{S_and_P}) of the two scalars that can be constructed from a $2$-form field strength and its Hodge dual in four spacetime dimensions. Similarly, let us define
\begin{align}
    \widetilde{H}^{a b c} = \frac{1}{3!} \epsilon^{a b c d e f} H_{def} \, ,
\end{align}
where $\epsilon^{abcdef}$ is the Levi-Civita symbol in six dimensions, and again indices are raised or lowered with the Euclidean metric $\delta_{ab}$ in this subsection. In this case, we will again find five independent invariants, which now occur at orders $2, 4, 4, 4, 6$, agreeing with \cite{Cederwall:2025ywy}.

To enumerate invariants using the algorithm of Section \ref{sec:algorithm}, we must now consider colored graphs, where the two possible colorings of a node correspond to $H_{abc}$ and $\widetilde{H}_{abc}$. As in the previous subsection, we must then enumerate all possible graphs for each fixed, even number of colored nodes, again with the property that every node has total degree $3$, to ensure that all indices are contracted. In practice, we accomplish this using a brute force approach: we use the graph-theoretic algorithm described in Section \ref{sec:trace} to first generate all \emph{uncolored} contraction graphs, then manually build all $2^{N}$ colorings of the $N$ nodes, and finally deduplicate redundancies using the isomorphism testing routine in igraph.

As before, non-trivial contraction graphs occur only for even $N$. At $N = 2$, there are three possible tensor networks,
\begin{center}
    \includegraphics[width=\linewidth]{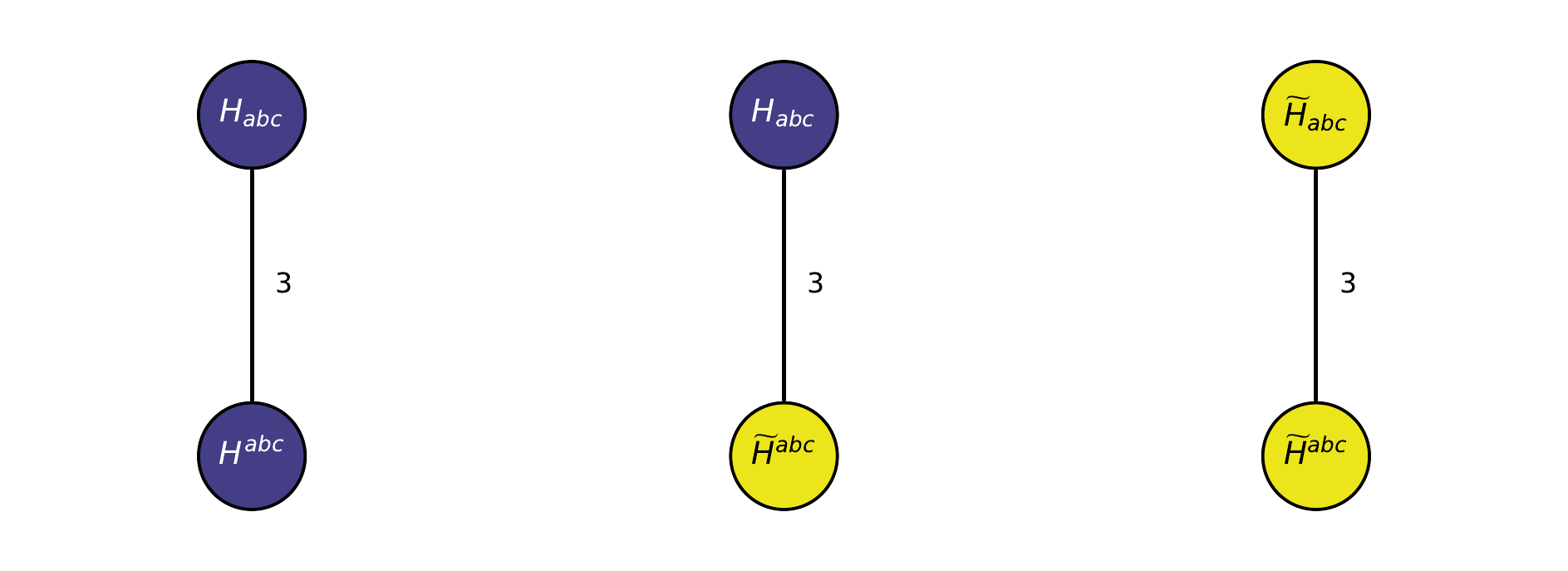}
\end{center}
However, $H_{abc} \widetilde{H}^{abc}$ vanishes identically, while $H_{abc} H^{abc}$ is proportional to $\widetilde{H}_{abc} \widetilde{H}^{abc}$, so there is only one independent invariant at this order, which we choose to be
\begin{align}
    y^{(2)} = H_{abc} H^{abc} = x^{(2)} \, .
\end{align}
At order $N = 4$, we find $12$ distinct colored contraction graphs, of which $4$ are linearly independent of one another, and only $3$ of these are also independent of $\left( y^{(2)} \right)^2$. The three new invariants at this order are represented by the graphs
\begin{center}
    \includegraphics[width=0.8\linewidth]{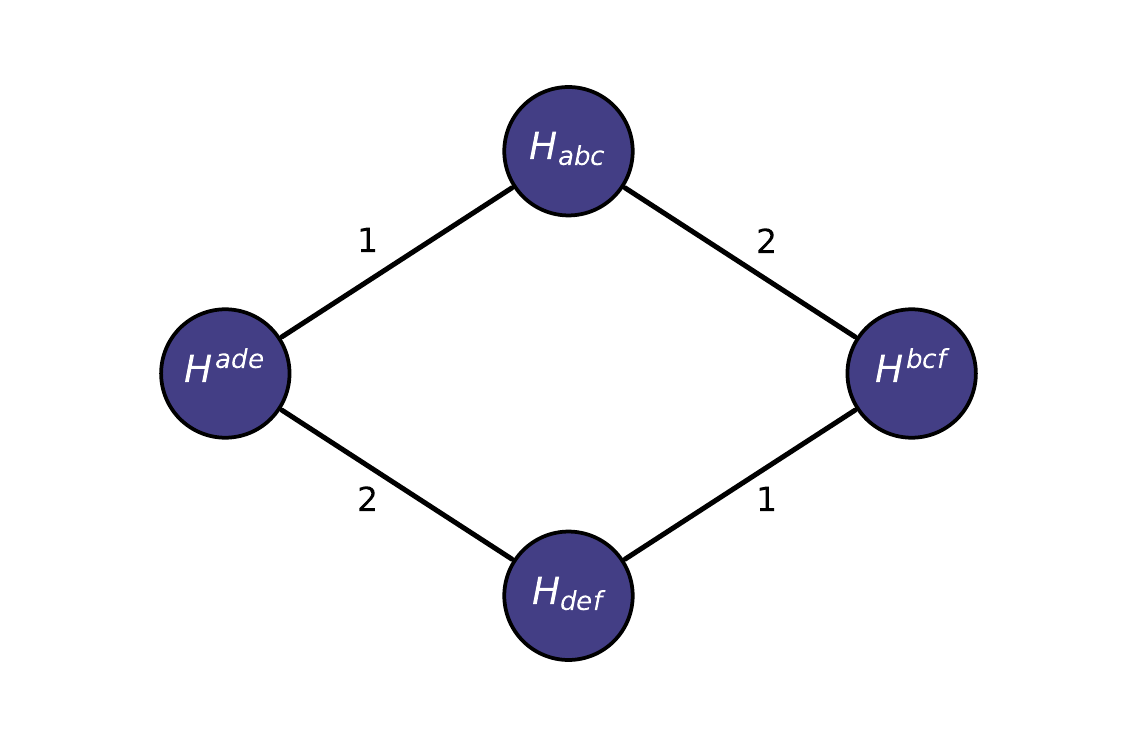}
\end{center}
and
\begin{center}
    \includegraphics[width=0.8\linewidth]{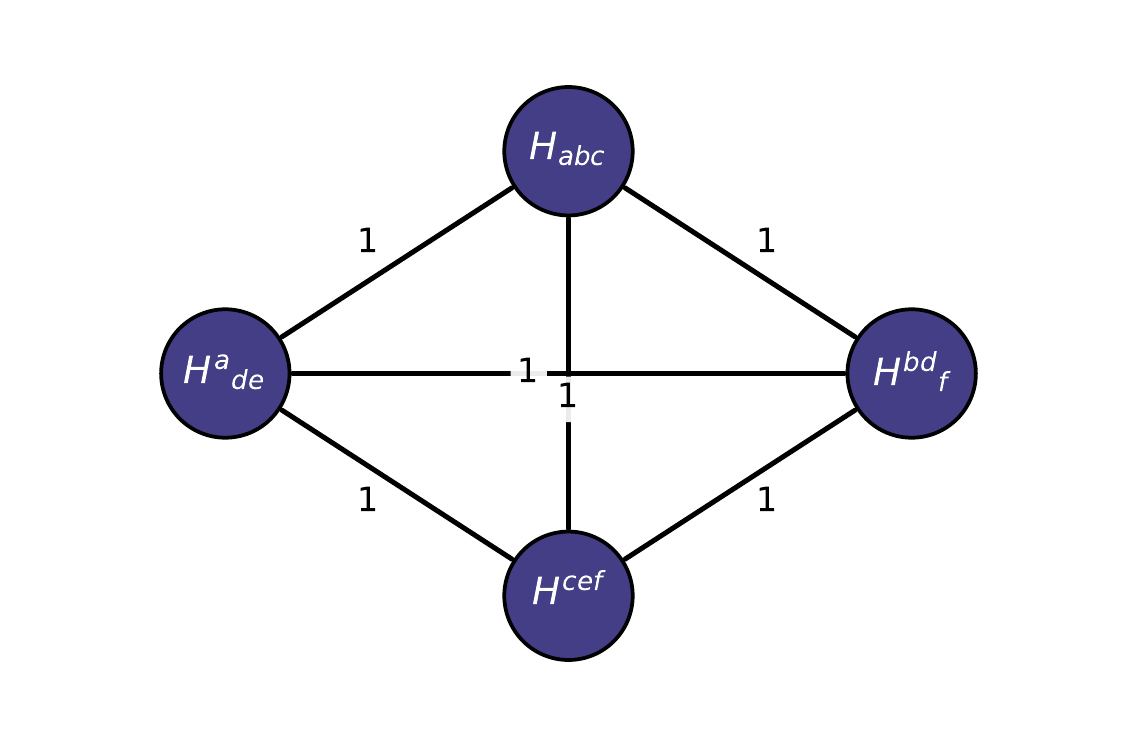}
\end{center}
and
\begin{center}
    \includegraphics[width=0.8\linewidth]{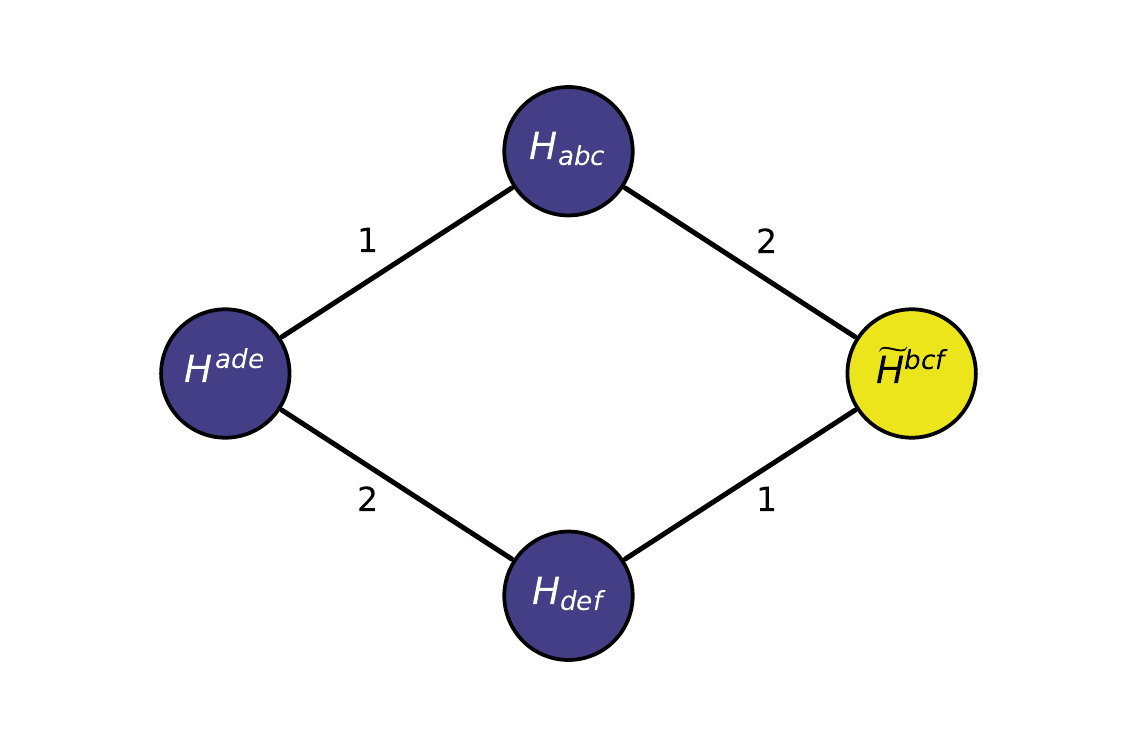}
\end{center}
which are associated with the invariants
\begin{align}
    y^{(4)}_1 &= H_{abc} H^{bcf} H_{def} H^{ade} = x_1^{(4)} \, , \nonumber \\
    y^{(4)}_2 &= H_{abc} \tensor{H}{^b^d_f} H^{cef} \tensor{H}{^a_d_e} = x_2^{(4)} \, , \nonumber \\
    y_3^{(4)} &= H_{abc} \widetilde{H}^{bcf} H_{def} H^{ade} \, .
\end{align}
We see that the first two invariants that the algorithm finds at this order are identical to those that were found in the trace variables algorithm of Section \ref{sec:trace}, which is a consistency check. However, the third invariant $y_3^{(4)}$ is new, and was ``missed'' in the preceding subsection because in that case we forbade the algorithm from using the invariant tensor $\epsilon^{abcdef}$, instead only allowing contractions involving the Kronecker delta. Therefore, using Hodge dual variables (in which both the Levi-Civita symbol and the Kronecker delta are employed as invariant tensors), we correctly find that there are three invariants at order $4$, as predicted from the generating function approach of \cite{Cederwall:2025ywy}.

Finally, at order $N = 6$, we find $114$ valid contraction graphs, of which only $5$ are linearly independent of one another, and only $1$ of these $5$ is independent of products of lower-order invariants. We conclude that there is only one novel invariant at sixth order, which can again be chosen to agree with the $6$th order invariant that we found in trace variables,
\begin{align}
    y^{(6)} = H_{abc} H^{chi} H_{ghi} H^{adg} H_{def} H^{bef} = x^{(6)} \, .
\end{align}
At this point, the algorithm terminates as expected, and all contraction graphs at orders $N \geq 8$ that we have considered are related to products of lower-order invariants.

We have therefore found a second conjectural set of independent and complete invariants that can be constructed from a $3$-form in $6d$, complementing the set found in Section \ref{sec:trace}. Given two such homogeneous systems of parameters, it is natural to wonder how to convert between them. For instance, in $4d$ Euclidean signature, one can translate\footnote{The fact that this map only involves the square $P^2$ shows that the trace variables can capture only parity-even combinations, which we have been cavalier about in this discussion.} between the two sets of variables (\ref{x1_and_x2}) and (\ref{S_and_P}) associated with a $2$-form field strength using the map
\begin{align}
    4 S - x_1 = 0 \, , \qquad  4 P^2 + x_2 - \frac{1}{2} x_1^2 = 0 \, .
\end{align}
In the present setting, most of the $y$ variables coincide with corresponding $x$ variables, except for the single invariant $y_3^{(4)}$ which is constructed using the Hodge dual. Because we have conjectured that the trace variables are a complete generating set, it should therefore follow that $y_3^{(4)}$ satisfies some polynomial relation involving the $x$ variables. Indeed, we do find such a relation using our algorithm: the invariants satisfy
\begin{align}
    \hspace{-5pt} 0 &= \left( y_3^{(4)} \right)^2 - 18 x^{(8)} - 8 x^{(6)} x^{(2)} - 6 x_1^{(4)} x_2^{(4)} + 4 \left( x_2^{(4)} \right)^2 + \left( x^{(2)} \right)^2 \left( \frac{2}{3} x_2^{(4)} + \frac{7}{6} x_1^{(4)} - \frac{1}{18} \left( x^{(2)} \right)^2 \right) \, ,
\end{align}
to within ten decimal places of numerical precision. This both gives further evidence that the collection of trace invariants of Section \ref{sec:trace} is complete, and provides us with a new identity which does not appear obvious from the analytical expressions for these invariants, but which is straightforward to obtain numerically.

\subsection{Spinor Variables}\label{sec:spinor}

A third way to describe the five independent invariants associated with a $3$-form in $6d$ is to use spinor indices, following the analogous definitions (\ref{spinor_variables}) for a $2$-form in four dimensions. For concreteness, we will review how this approach proceeds using Lorentz signature and the isomorphism $\mathfrak{so} ( 1, 5 ) \cong \mathfrak{su}^\ast ( 4 )$.\footnote{We refer the reader to the papers \cite{Linch:2012zh,Butter:2016qkx,Kennedy:2025nzm} for further details; in each of these three works, the respective Appendix A contains a very useful reference on six-dimensional spinorial conventions and this isomorphism.} Consider the $8 \times 8$ gamma matrices $\Gamma^\mu$ which obey
\begin{align}\label{gamma_conventions}
    \left\{ \Gamma_\mu , \Gamma_\nu \right\} = - 2 \eta_{\mu \nu} \mathbb{I} \, .
\end{align}
One can take these matrices to be block off-diagonal with the form
\begin{align}
    \Gamma^\mu = \begin{bmatrix} 0 & \left( \tilde{\gamma}^\mu \right)^{\alpha \beta} \\ \left( \gamma^\mu \right)_{\alpha \beta} & 0 \end{bmatrix} \, ,
\end{align}
where now early Greek indices like $\alpha, \beta$ run from $1$ to $4$. For each $\mu$, the matrices $\left( \gamma^\mu \right)_{\alpha \beta}$ and $\left( \tilde{\gamma}^\mu \right)^{\alpha \beta}$ are antisymmetric in their spinor indices and satisfy
\begin{align}
    \left( \tilde{\gamma}^\mu \right)^{\alpha \beta} = \frac{1}{2} \epsilon^{\alpha \beta \gamma \delta} \left( \gamma^\mu \right)_{\gamma \delta} \, .
\end{align}
From these objects, we can define products of three such matrices by
\begin{align}
    \gamma_{\mu \nu \rho} = \gamma_{[\mu} \tilde{\gamma}_\nu \gamma_{\rho]} \, , \qquad \tilde{\gamma}_{\mu \nu \rho} = \tilde{\gamma}_{[\mu} \gamma_\nu \tilde{\gamma}_{\rho]} \, .
\end{align}
Now we turn to three-forms. The $20$-dimensional $3$-form representation of $SO(1, 5)$ is not irreducible, but rather splits into the two irreducible representations of self-dual and anti-self-dual tensors. That is, any $3$-form $H_{\mu \nu \rho}$ can be written as
\begin{align}
    H_{\mu \nu \rho} = H_{\mu \nu \rho}^{+} + H_{\mu \nu \rho}^{-} \, ,
\end{align}
where
\begin{align}
    \frac{1}{3!} \epsilon^{\mu \nu \rho \sigma \kappa \lambda} H_{\sigma \kappa \lambda}^{\pm} = \pm H^{\pm \mu \nu \rho} \, .
\end{align}
By contracting spacetime indices, we may therefore construct two symmetric $4 \times 4$ matrices $M^{\alpha \beta}$ and $N_{\alpha \beta}$ from a generic $3$-form $H_{\mu \nu \rho}$ as
\begin{align}
    M^{\alpha \beta} &= \frac{1}{3!} \left( \tilde{\gamma}^{\mu \nu \rho} \right)^{\alpha \beta} H_{\mu \nu \rho} \, , \nonumber \\
    N_{\alpha \beta} &= \frac{1}{3!} \left( \gamma^{\mu \nu \rho} \right)_{\alpha \beta} H_{\mu \nu \rho} \, ,
\end{align}
from which one can recover the self-dual and anti-self-dual parts of $H_{\mu \nu \rho}$, respectively, as
\begin{align}\label{go_back_to_3form}
    H^+_{\mu \nu \rho} &= \frac{1}{8} \left( \tilde{\gamma}_{\mu \nu \rho} \right)^{\alpha \beta} N_{\alpha \beta} \, ,  \nonumber \\
    H^-_{\mu \nu \rho} &= \frac{1}{8} \left( \gamma_{\mu \nu \rho} \right)_{\alpha \beta} M^{\alpha \beta} \, .
\end{align}
Thus the problem of enumerating the invariants that can be constructed from a three-form in six dimensions is in one-to-one correspondence with the problem of enumerating such invariants for a pair of $4 \times 4$ matrices $M^{\alpha \beta}$ and $N_{\alpha \beta}$.\footnote{We thank Gabriele Tartaglino-Mazzucchelli and Dmitri Sorokin for helpful discussions on this point.} One caveat is that the spinor indices on these objects cannot be raised or lowered with a metric, so the only allowed contractions are those between an upper and lower index either of the matrices $M^{\alpha \beta}$ and $N_{\alpha \beta}$, or involving the antisymmetric symbol $\epsilon^{\alpha \beta \gamma \delta}$ or $\epsilon_{\alpha \beta \gamma \delta}$ of $\mathfrak{su}^\ast ( 4 )$.

As an aside, we note that spinor variables make it quite transparent that, for a \emph{chiral} $3$-form (self-dual or anti-self-dual) in six dimensions, there is only one independent invariant that can be constructed. For the self-dual (anti-self-dual) case, the matrix $M^{\alpha \beta}$ ($N_{\alpha \beta}$) vanishes, and the invariant theory problem reduces to the enumeration of invariants that can be built from a single $4 \times 4$ matrix and the Levi-Civita symbol. Clearly the only such invariant available is the determinant, which is quartic in the matrix. This corresponds to the single quartic invariant which appeared as a singlet in the tensor product of four self-dual or anti-self-dual tensor representations of $SO(6)$ in the LieART output presented below equation (\ref{molien_weyl}) in Section \ref{sec:review}. This fact was used in \cite{Avetisyan:2022zza} which described a new formulation of Lagrangians for self-dual tensors in diverse dimensions, including a chiral $3$-form in $6d$, which can also be understood using the PST formalism \cite{Pasti:1995ii,Pasti:1995tn,Pasti:1996vs}. However, we are presently interested in \emph{non-chiral} $3$-forms, for which both matrices $M^{\alpha \beta}$ and $N_{\alpha \beta}$ are non-vanishing.

We will now address this equivalent two-matrix invariant theory problem using our main algorithm from Section \ref{sec:algorithm}. We again find five total invariants at total orders $2, 4, 4, 4, 6$ in the matrices $M^{\alpha \beta}$ and $N_{\alpha \beta}$. Here the graph theoretic algorithm must again be modified slightly: we will enumerate all directed weighted graphs that can be constructed using the nodes $M^{\alpha \beta}$, $N_{\alpha \beta}$, $\epsilon^{\alpha \beta \gamma \delta}$, and $\epsilon_{\alpha \beta \gamma \delta}$. The direction of an edge indicates contraction between an upper index on the source and a lower index on the sink. We demand that each $M^{\alpha \beta}$ node has out-degree exactly equal to $2$ (and no inward-pointing edges), each $N_{\alpha \beta}$ node has in-degree equal to $2$ (and no outward-pointing edges), and similarly that the Levi-Civita symbols have total out-degree or in-degree equal to $4$ for $\epsilon^{\alpha \beta \gamma \delta}$ or $\epsilon_{\alpha \beta \gamma \delta}$, respectively. We construct all allowed contraction graphs using an exact backtracking enumeration of the same form that was described in Section \ref{sec:trace}. Because $M^{\alpha \beta}$ and $N_{\alpha \beta}$ are symmetric in their indices, while the Levi-Civita tensors are antisymmetric, this approach suffices since we need not track which precise index is contracted; any contraction on a particular index is related to a contraction on a different index by symmetry. As before, we deduplicate isomorphic graphs by testing for graph isomorphisms and retaining only one representative of each isomorphism class.

At order $N = 2$, there is a unique contraction graph
\begin{center}
    \includegraphics[width=0.45\linewidth]{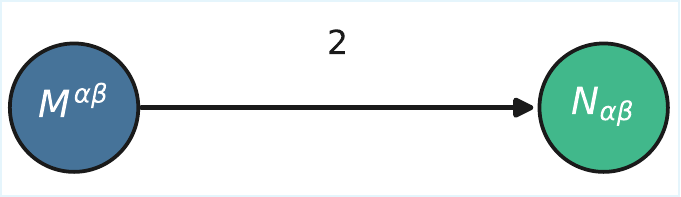}
\end{center}
corresponding to the invariant
\begin{align}
    z^{(2)} = M^{\alpha \beta} N_{\alpha \beta} \, .
\end{align}
At order $N = 4$, after performing the steps of our algorithm, we find three independent invariants, corresponding to the contraction graphs
\begin{center}
    \includegraphics[width=0.6\linewidth]{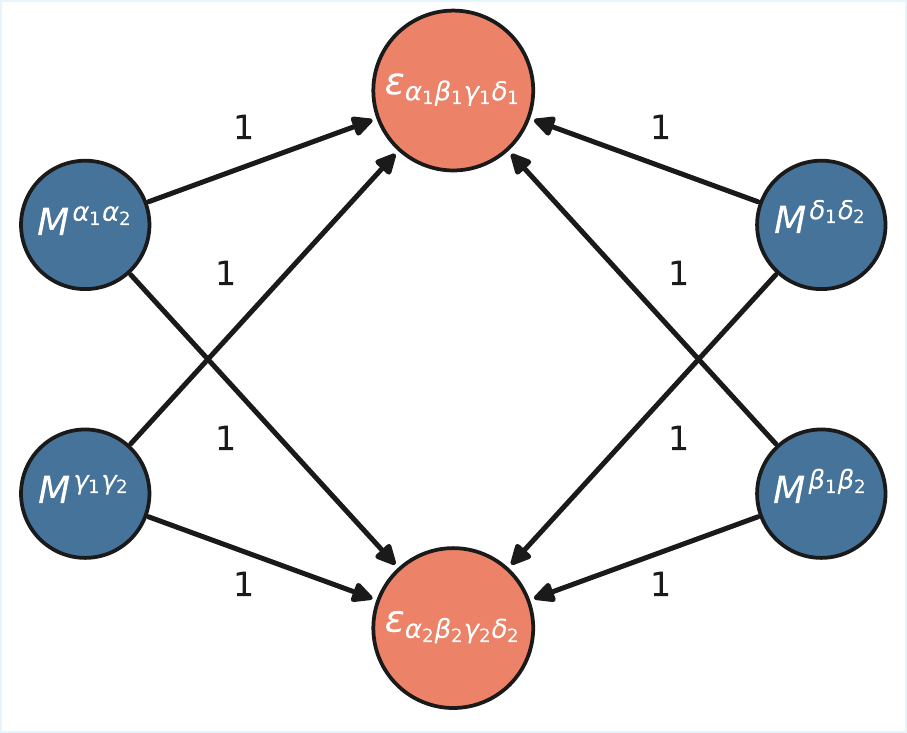}
\end{center}
and
\begin{center}
    \includegraphics[width=0.6\linewidth]{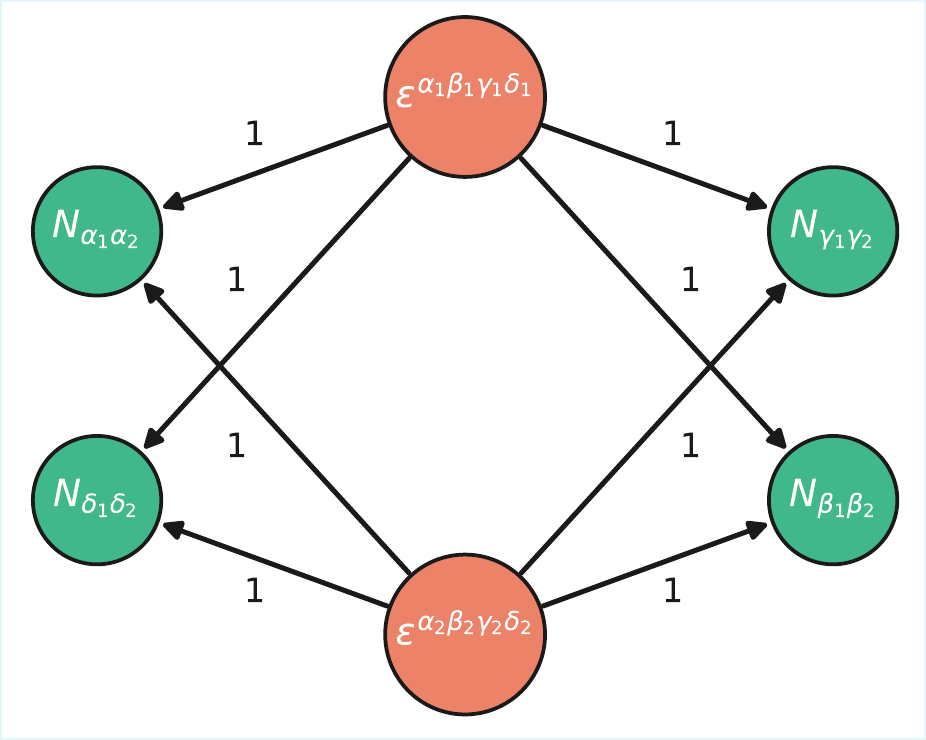}
\end{center}
and
\begin{center}
    \includegraphics[width=0.6\linewidth]{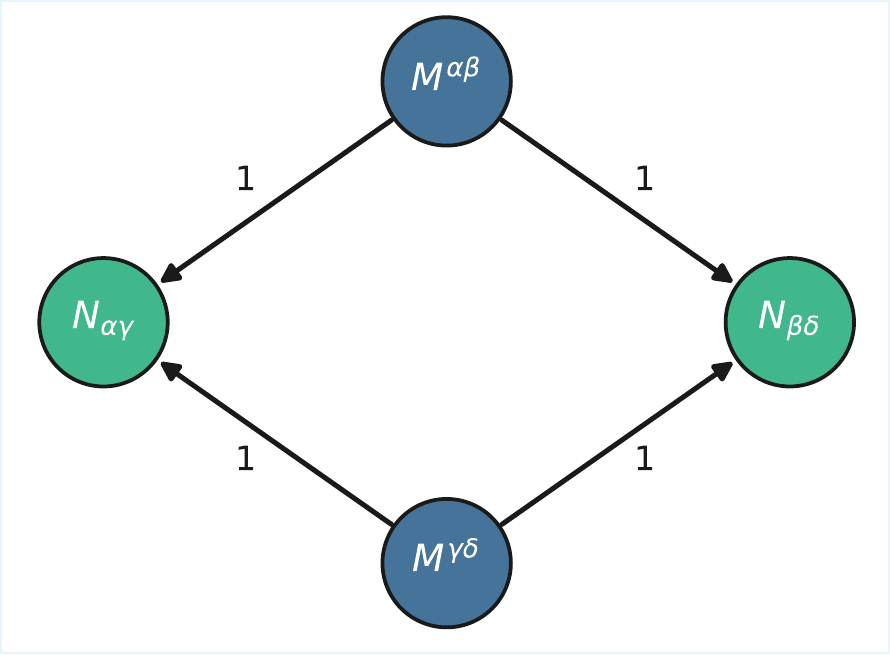}
\end{center}
Two of these are simply proportional to the determinants of the matrices $M^{\alpha \beta}$ and $N_{\alpha \beta}$,
\begin{align}
    z^{(4)}_1 &= \epsilon_{\alpha_1 \beta_1 \gamma_1 \delta_1} \epsilon_{\alpha_2 \beta_2 \gamma_2 \delta_2} M^{\alpha_1 \alpha_2} M^{\beta_1 \beta_2} M^{\gamma_1 \gamma_2} M^{\delta_1 \delta_2} \, , \nonumber \\
    z^{(4)}_2 &= \epsilon^{\alpha_1 \beta_1 \gamma_1 \delta_1} \epsilon^{\alpha_2 \beta_2 \gamma_2 \delta_2} N_{\alpha_1 \alpha_2} N_{\beta_1 \beta_2} N_{\gamma_1 \gamma_2} N_{\delta_1 \delta_2} \, ,
\end{align}
while the third invariant corresponds to the contraction
\begin{align}
    z^{(4)}_3 = M^{\alpha \beta} M^{\gamma \delta} N_{\alpha \gamma} N_{\beta \delta} \, .
\end{align}
At order $N = 6$, we find one independent invariant, represented by the graph
\begin{center}
    \includegraphics[width=0.6\linewidth]{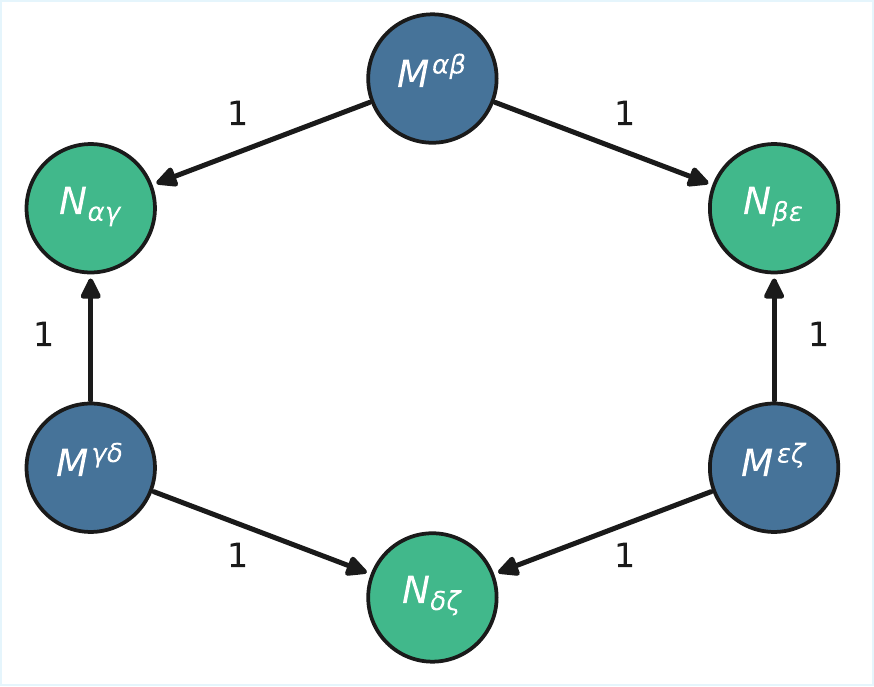}
\end{center}
and corresponding to the contraction
\begin{align}
    z^{(6)} = M^{\alpha \beta} M^{\gamma \delta} M^{\epsilon \zeta} N_{\alpha \gamma} N_{\beta \epsilon} N_{\delta \zeta} \, .
\end{align}
Now our algorithm again terminates, and we find that all invariants associated with contraction graphs for $N \geq 8$ can be expressed in terms of lower-order invariants.

Let us make one additional numerological remark. If we again count the total number of contraction graphs appearing at each order in $N$, rather than only those corresponding to independent invariants, we find the counts $1, 4, 11, 22, 38$ for $N = 2, 4, 6, 8, 10$, respectively. These agree with the first terms in several coordination sequences for certain zeolite compounds, such as \cite{zeoliteOne} and \cite{zeoliteTwo}. This formal similarity appears to be due to the fact that such zeolites are constructed from silicon atoms along with other atoms, such as oxygen, and silicon forms four covalent bonds while oxygen forms two. This mirrors the structure of the contraction graphs that we are constructing, since e.g. Levi-Civita symbols require four index contractions (resembling the four bonds formed by silicon) while the matrices $M^{\alpha \beta}$ and $N_{\alpha \beta}$ require two index contractions (like the two bonds formed by oxygen).

To conclude, we have given an explicit set of invariants that can be constructed from two $4 \times 4$ symmetric matrices $M^{\alpha \beta}$ and $N_{\alpha \beta}$, which we conjecture to be independent and complete based on substantial numerical evidence. Converting these objects back to the self-dual and anti-self-dual parts $H^{\pm}_{\mu \nu \rho}$ using (\ref{go_back_to_3form}), and summing to recover $H_{\mu \nu \rho}$, therefore gives a corresponding set of invariants that can be constructed from a $3$-form in six dimensions. We recall that we are using the conventions (\ref{gamma_conventions}) for gamma matrices when performing this conversion, as appropriate for Lorentz signature, so it is not possible to directly give a map between the invariants constructed in this subsection and those in Sections \ref{sec:trace} and \ref{sec:hodge}, as both of those analyses used Euclidean signature. The advantage of using Lorentz signature for the present analysis is that it experimentally confirms the natural expectation that the number of independent invariants is independent of spacetime signature. Furthermore, one can always obtain explicit expressions for the invariants constructed here in terms of contractions of $H_{\mu \nu \rho}$ with the Minkowski metric $\eta_{\mu \nu}$ using various identities, such as
\begin{align}
    \left( \gamma^{\mu \nu \rho} \right)_{\alpha \beta} \left( \tilde{\gamma}_{\mu \nu \rho} \right)^{\gamma \delta} = 48 \tensor{\delta}{_{(\alpha}^{\gamma}} \tensor{\delta}{_{\beta)}^{\delta}} \, ,
\end{align}
which holds in our conventions.

\section{Conclusion}\label{sec:conclusion}

In this work, we described a numerical approach to experimentally studying the independent invariants that can be built from tensors. We have applied this technique to enumerate the scalars that can be constructed from a $3$-form tensor field in six dimensions, using three different sets of variables, and consistently find a set of five independent invariants in each case. We also provided a dictionary that allows conversion between two of these sets of variables, and give some examples of relations between other invariants and those in the independent generating set. These explicit expressions might be used as a starting point for studying properties of the family of $3$-form theories in $6d$, much as the parameterizations  (\ref{x1_and_x2}), (\ref{S_and_P}), (\ref{spinor_variables}) can be used to study general theories of non-linear electrodynamics in $4d$.

There remain several intriguing directions for future research. Perhaps the most obvious is to use this data-driven method to study the scalars that can be constructed from other tensors and in different numbers of dimensions. For instance, it would be interesting to numerically study the invariants that are associated with a chiral $5$-form in ten dimensions, which were recently considered in \cite{Cederwall:2025ywy,Hutomo:2025dfx} and in particular used to formulate ModMax-type theories \cite{Bandos:2020jsw} in ten dimensions. One might also investigate invariants of interest for effective actions (like those constructed from the Riemann tensor in ten dimensions), to local unitary\footnote{See \cite{Ferko:2025jrf} for recent work introducing new LU invariants based on the cohomological formalism of \cite{Mainiero:2019enr,Ferko:2024swt}.} and SLOCC invariants, and to multi-matrix models, corresponding to the three motivating examples \ref{stringy}, \ref{quantum}, and \ref{multi_matrix} that we mentioned in Section \ref{sec:intro}.

Another advantage of this formalism is that it generates the necessary trace identities which are needed to simplify classical $\TT$-like flow equations \cite{Zamolodchikov:2004ce,Smirnov:2016lqw,Cavaglia:2016oda}, and derive the corresponding master equations, for deformations of tensor fields. Although the main results of this work concern $3$-forms, let us work more generally for the moment. Beginning from a free $p$-form field in $d$ dimensions, whose action (up to normalization) is 
\begin{align}\label{free_p_form}
    S = \int d^d x \, \sqrt{-g} \, H_{\mu_1 \cdots \mu_p} H^{\mu_1 \cdots \mu_p} \, ,
\end{align}
the Hilbert stress tensor is defined by taking a functional derivative with respect to $g_{\mu \nu}$, so
\begin{align}
    T_{\mu \nu} &= - 2 \frac{\partial \mathcal{L}}{\partial g^{\mu \nu}} + g_{\mu \nu} \mathcal{L} \nonumber \\
    &= - 2 p H_{\mu \mu_2 \cdots \mu_p} \tensor{H}{_\nu^{\mu_2}^{\cdots}^{\mu_p}} + g_{\mu \nu} H_{\mu_1 \cdots \mu_p} H^{\mu_1 \cdots \mu_p} \, .
\end{align}
Defining $|H|^2 = H_{\mu_1 \cdots \mu_p} H^{\mu_1 \cdots \mu_p}$ and $|H|^2_{\mu \nu} =  H_{\mu \mu_2 \cdots \mu_p} \tensor{H}{_\nu^{\mu_2}^{\cdots}^{\mu_p}}$, the stress tensor is
\begin{align}
    T_{\mu \nu} = - 2 p | H |^2_{\mu \nu} + g_{\mu \nu} | H |^2 \, .
\end{align}
Now suppose that one wishes to deform the classical action by an integrated function of the energy-momentum tensor,
\begin{align}
    \frac{\partial S}{\partial \lambda} = \int d^d x \, \sqrt{-g} \, f ( T_{\mu \nu} ) \, .
\end{align}
For instance, we can allow the function $f$ to depend on the two Lorentz invariants $\tensor{T}{^\mu_\mu}$ and $T_{\mu \nu} T^{\mu \nu}$, which for the free theory are given by (\ref{free_p_form}), finding
\begin{align}
    \tensor{T}{^\mu_\mu} &= \left( d - 2 p \right) | H |^2 \, , \nonumber \\
    T_{\mu \nu} T^{\mu \nu} &= 4 p^2 | H |^2_{\mu \nu} | H |^{2 \mu \nu} + ( d - 4 p ) | H |^4 \, .
\end{align}
As one continues to iterate along a $\TT$-like flow, more and more invariants built from contractions of $H$ are generated, and eventually one must use identities to relate the new scalars to those in a minimal generating set. For instance, in the case of a $3$-form in six dimensions, one could use the results of this work to derive a partial differential equation in six variables (the five independent invariants associated with $H_{\mu \nu \rho}$, and the flow parameter $\lambda$) which represents a ``master flow equation'' for classical $\TT$-like deformations.\footnote{For other results on solving classical $\TT$-like flows, see the incomplete sampling of works \cite{Bonelli:2018kik,Hou:2022csf,Ferko:2022dpg,Ferko:2023sps,Blair:2024aqz,Babaei-Aghbolagh:2024uqp}, and for general information about the $\TT$ deformation, we refer the reader to the reviews \cite{Jiang:2019epa,He:2025ppz}.} It would be interesting to obtain such a master PDE and see whether any particular deforming operators give rise to deformed theories with special properties, as has been investigated for $2$-forms in $4d$ \cite{Conti:2018jho,Ferko:2021loo,Ferko:2022iru}, for chiral bosons in $2d$ \cite{Ebert:2024zwv}, and for chiral $3$-forms in $6d$ \cite{Ferko:2024zth}. Another direction would be to apply the technique of this article to enumerating invariants that can be constructed from a non-Abelian field strength with components $F_{\mu \nu} = F_{\mu \nu}^A T_A$, where $T_A$ are generators for the Lie algebra $\mathfrak{g}$ of the gauge group, and use this to continue the study of classical $\TT$-like deformations of Yang-Mills theories in $d = 4$, initiated in \cite{Ferko:2024yua}.

We hope to return to some of these interesting questions in future work.

\section*{Acknowledgements}

We are very grateful to Ning Bao, James Halverson, Dmitri Sorokin, Gabriele Tartaglino-Mazzucchelli, and Linus Wulff for helpful discussions about the subject of this work. C.\,F. is supported by the National Science Foundation under Cooperative Agreement PHY-2019786 (the NSF AI Institute for Artificial Intelligence and Fundamental Interactions).


\appendix

\section{Additional Graphs and Relations}\label{app:relations}

In this Appendix, we collect a few additional contraction graphs and relations which were omitted in the body of this article for brevity. The primary purpose of this exercise is to demonstrate how our algorithm can generate additional non-trivial relations between various invariants in some examples.

In Section \ref{sec:trace}, we commented that only one genuine novel invariant is introduced at order $N = 6$. Let us consider the other two graphs which were presented in that section at this order, corresponding to the contractions
\begin{align}
    X_1^{(6)} &= H_{abc} H^{bci} H_{ghi} H^{fgh} H_{def} H^{ade} \, , \nonumber \\
    X_2^{(6)} &= H_{abc} H^{cfh} H_{def} H^{bei} \tensor{H}{^d^g_i} \tensor{H}{^a_g_h} \, .
\end{align}
We use capital letters $X$ for these invariants to distinguish them from those in the independent generating set, such as $x^{(6)}$. These invariants should satisfy polynomial relations which allow them to be expressed in terms of the $x$ variables, and numerically, we find
\begin{align}
    0 &= X_1^{(6)} - \frac{1}{2} x^{(2)} x_1^{(4)} + \frac{1}{18} \left( x^{(2)} \right)^3 \, , \nonumber \\
    0 &= X_2^{(6)} + \frac{1}{2} x^{(6)} + \frac{1}{12} x^{(2)} x_2^{(4)} - \frac{1}{6} x^{(2)} x_1^{(4)} + \frac{1}{72} \left( x^{(2)} \right)^3 \, . \label{app_id_1}
\end{align}
At order $N = 8$, we commented that one finds $20$ non-isomorphic connected contraction graphs, of which only $6$ are linearly independent of the others, but that only one of these $6$ is also independent of products of lower-order invariants. Let us consider a couple of the dependent graphs at this order. One is the tensor network
\begin{center}
    \includegraphics[width=0.7\linewidth]{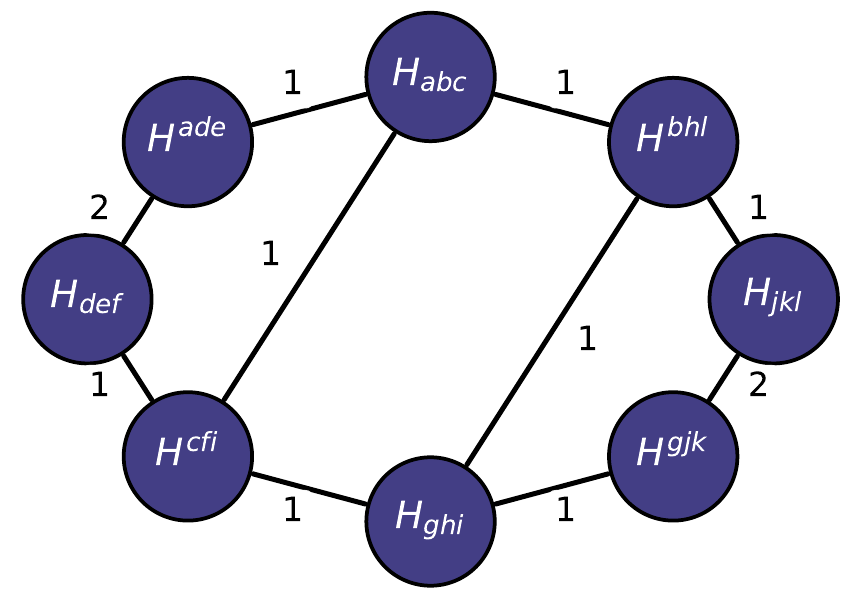}
\end{center}
associated with the invariant
\begin{align}
    X_1^{(8)} = H_{abc} H^{bhl} H_{jkl} H^{gjk} H_{ghi} H^{cfi} H_{def} H^{ade} \, .
\end{align}
Again, this quantity satisfies a polynomial relation involving the generators, namely
\begin{align}\label{app_id_2}
    0 &= X_1^{(8)} + \frac{5}{2} x^{(8)} + \frac{3}{2} x^{(6)} x^{(2)} + x^{(4)}_1 x^{(4)}_2 - \frac{2}{3} \left( x^{(4)}_2 \right)^2 + \frac{1}{4} \left( x_1^{(4)} \right)^2 - \frac{1}{9} x^{(4)}_2 \left( x^{(2)} \right)^2 \nonumber \\
    &\qquad \qquad - \frac{11}{36} x^{(4)}_1 \left( x^{(2)} \right)^2 + \frac{1}{54} \left( x^{(2)} \right)^4 \, .
\end{align}
Another dependent graph at eighth order is
\begin{center}
    \includegraphics[width=0.7\linewidth]{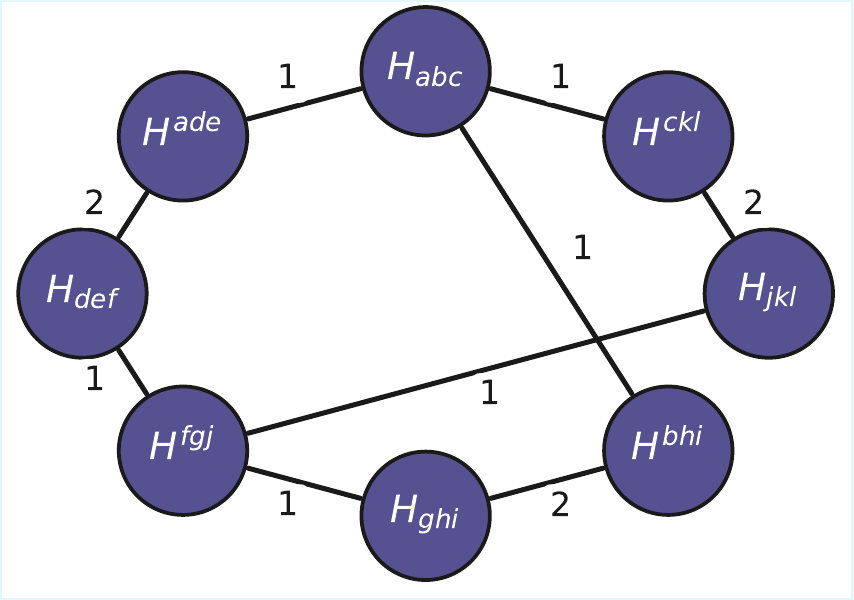}
\end{center}
whose associated invariant is
\begin{align}
    X_2^{(8)} = H_{abc} H^{ckl} H_{jkl} H^{bhi} H_{ghi} H^{fgj} H_{def} H^{ade} \, ,
\end{align}
which satisfies the polynomial relation
\begin{align}\label{app_id_3}
    0 &= X_2^{(8)} + 3 x^{(8)} + \frac{2}{3} x^{(6)} x^{(2)} + x^{(4)}_1 x^{(4)}_2 - \frac{2}{3} \left( x_2^{(4)} \right)^2 - \frac{1}{9} x_2^{(4)} \left( x^{(2)} \right)^2 - \frac{1}{18} x_1^{(4)} \left( x^{(2)} \right)^2 \, .
\end{align}
Clearly this procedure can be repeated to express any dependent invariant as a polynomial in the generators. Operationally, in each case, we first obtain a numerical approximation to the null vector of the data matrix which is obtained by drawing many random values of the tensor $H_{abc}$ and computing the contractions corresponding to each invariant, as described in Section \ref{sec:algorithm}. We then obtain a rational approximation to these numerical values using the \texttt{Rationalize} function in Mathematica. Finally, we have tested the candidate identities like (\ref{app_id_1}), (\ref{app_id_2}), (\ref{app_id_3}) by drawing new random values for the tensor components $1,000$ times each, computing the combination of contractions appearing on the right side of these equations, and verifying that the sum vanishes to within numerical precision. 
 
\bibliographystyle{utphys}
\bibliography{master}

@article{Cederwall:2025ywy,
    author = "Cederwall, Martin and Hutomo, Jessica and Kuzenko, Sergei M. and Lechner, Kurt and Sorokin, Dmitri P.",
    title = "{Some remarks on invariants}",
    eprint = "2509.14350",
    archivePrefix = "arXiv",
    primaryClass = "hep-th",
    month = "9",
    year = "2025"
}

@article{Pasti:1996vs,
    author = "Pasti, Paolo and Sorokin, Dmitri P. and Tonin, Mario",
    title = "{On Lorentz invariant actions for chiral p forms}",
    eprint = "hep-th/9611100",
    archivePrefix = "arXiv",
    reportNumber = "DFPD-96-TH-58",
    doi = "10.1103/PhysRevD.55.6292",
    journal = "Phys. Rev. D",
    volume = "55",
    pages = "6292--6298",
    year = "1997"
}

@article{Pasti:1995tn,
    author = "Pasti, Paolo and Sorokin, Dmitri P. and Tonin, Mario",
    title = "{Duality symmetric actions with manifest space-time symmetries}",
    eprint = "hep-th/9506109",
    archivePrefix = "arXiv",
    reportNumber = "DFPD-95-TH-37, PREPRINT-DFPD-95-TH-37",
    doi = "10.1103/PhysRevD.52.R4277",
    journal = "Phys. Rev. D",
    volume = "52",
    pages = "R4277--R4281",
    year = "1995"
}

@phdthesis{Ferko:2021loo,
    author = "Ferko, Christian",
    title = "{Supersymmetry and Irrelevant Deformations}",
    eprint = "2112.14647",
    archivePrefix = "arXiv",
    primaryClass = "hep-th",
    school = "Chicago U.",
    year = "2021"
}

@article{Babaei-Aghbolagh:2022uij,
    author = "Babaei-Aghbolagh, H. and Velni, Komeil Babaei and Yekta, Davood Mahdavian and Mohammadzadeh, H.",
    title = "{Emergence of non-linear electrodynamic theories from $T\bar{T}$-like deformations}",
    eprint = "2202.11156",
    archivePrefix = "arXiv",
    primaryClass = "hep-th",
    month = "2",
    year = "2022"
}

@book{Olver_1999, place={Cambridge}, series={London Mathematical Society Student Texts}, title={Classical Invariant Theory}, publisher={Cambridge University Press}, author={Olver, Peter J.}, year={1999}, collection={London Mathematical Society Student Texts}}

@article{Feger:2012bs,
    author = "Feger, Robert and Kephart, Thomas W.",
    title = "{LieART{\textemdash}A Mathematica application for Lie algebras and representation theory}",
    eprint = "1206.6379",
    archivePrefix = "arXiv",
    primaryClass = "math-ph",
    doi = "10.1016/j.cpc.2014.12.023",
    journal = "Comput. Phys. Commun.",
    volume = "192",
    pages = "166--195",
    year = "2015"
}

@article{HochsterRoberts1974,
  author = {Hochster, M. and Roberts, J. L.},
  title = {Rings of invariants of reductive groups acting on regular rings are Cohen-Macaulay},
  journal = {Advances in Mathematics},
  volume = {13},
  pages = {115--175},
  year = {1974}
}

@article{Feger:2019tvk,
    author = "Feger, Robert and Kephart, Thomas W. and Saskowski, Robert J.",
    title = "{LieART 2.0 {\textendash} A Mathematica application for Lie Algebras and Representation Theory}",
    eprint = "1912.10969",
    archivePrefix = "arXiv",
    primaryClass = "hep-th",
    doi = "10.1016/j.cpc.2020.107490",
    journal = "Comput. Phys. Commun.",
    volume = "257",
    pages = "107490",
    year = "2020"
}

@manual{vanLeeuwenCohenLisser_LiE_1992,
  title        = {{LiE, A Package for Lie Group Computations}},
  author       = {van Leeuwen, M. A. A. and Cohen, A. M. and Lisser, B.},
  organization = {{Computer Algebra Nederland}},
  address      = {Amsterdam},
  year         = {1992},
  isbn         = {90-74116-02-7},
}

@book{goodman2000representations,
  title={Representations and Invariants of the Classical Groups},
  author={Goodman, R. and Wallach, N.R.},
  isbn={9780521663489},
  lccn={97032151},
  series={Encyclopedia of Mathematics and its Applications},
  url={https://books.google.com/books?id=MYFepb2yq1wC},
  year={2000},
  publisher={Cambridge University Press}
}

@article{Bandos:2020jsw,
    author = "Bandos, Igor and Lechner, Kurt and Sorokin, Dmitri and Townsend, Paul K.",
    title = "{A non-linear duality-invariant conformal extension of Maxwell's equations}",
    eprint = "2007.09092",
    archivePrefix = "arXiv",
    primaryClass = "hep-th",
    doi = "10.1103/PhysRevD.102.121703",
    journal = "Phys. Rev. D",
    volume = "102",
    pages = "121703",
    year = "2020"
}

@article{He:2025ppz,
    author = "He, Song and Li, Yi and Ouyang, Hao and Sun, Yuan",
    title = "{$T\overline{T}$ Deformation: Introduction and Some Recent Advances}",
    eprint = "2503.09997",
    archivePrefix = "arXiv",
    primaryClass = "hep-th",
    month = "3",
    year = "2025"
}

@article{Blair:2024aqz,
    author = "Blair, Chris D. A. and Lahnsteiner, Johannes and Obers, Niels A. and Yan, Ziqi",
    title = "{Matrix theory reloaded: a BPS road to holography}",
    eprint = "2410.03591",
    archivePrefix = "arXiv",
    primaryClass = "hep-th",
    reportNumber = "IFT-UAM/CSIC-24-139, NORDITA 2024-033",
    doi = "10.1007/JHEP02(2025)024",
    journal = "JHEP",
    volume = "02",
    pages = "024",
    year = "2025"
}

@article{Hou:2022csf,
    author = "Hou, Jue",
    title = "{$ T\overline{T} $ flow as characteristic flows}",
    eprint = "2208.05391",
    archivePrefix = "arXiv",
    primaryClass = "hep-th",
    doi = "10.1007/JHEP03(2023)243",
    journal = "JHEP",
    volume = "03",
    pages = "243",
    year = "2023"
}

@article{Babaei-Aghbolagh:2024uqp,
    author = "Babaei-Aghbolagh, H. and He, Song and Ouyang, Hao",
    title = "{Generalized $ T\overline{T} $-like deformations in duality-invariant nonlinear electrodynamic theories}",
    eprint = "2407.03698",
    archivePrefix = "arXiv",
    primaryClass = "hep-th",
    doi = "10.1007/JHEP09(2024)137",
    journal = "JHEP",
    volume = "09",
    pages = "137",
    year = "2024"
}

@article{Jiang:2019epa,
    author = "Jiang, Yunfeng",
    title = "{A pedagogical review on solvable irrelevant deformations of 2D quantum field theory}",
    eprint = "1904.13376",
    archivePrefix = "arXiv",
    primaryClass = "hep-th",
    reportNumber = "CERN-TH-2019-058",
    doi = "10.1088/1572-9494/abe4c9",
    journal = "Commun. Theor. Phys.",
    volume = "73",
    number = "5",
    pages = "057201",
    year = "2021"
}

@article{Gibbons:1995cv,
    author = "Gibbons, G. W. and Rasheed, D. A.",
    title = "{Electric - magnetic duality rotations in nonlinear electrodynamics}",
    eprint = "hep-th/9506035",
    archivePrefix = "arXiv",
    doi = "10.1016/0550-3213(95)00409-L",
    journal = "Nucl. Phys. B",
    volume = "454",
    pages = "185--206",
    year = "1995"
}

@article{BialynickiBirula:1984tx,
      author         = "Bialynicki-Birula, I.",
      title          = "{Nonlinear Electrodynamics: variations on a theme by Born and Infeld}",
       journal        = "In: ``Quantum Theory Of Particles and Fields: Birthday Volume Dedicated to Jan Lopuszanski'' (Eds. B. Jancewicz and J. Lukierski), World Scientific Publishing Co Pte Ltd, 1984, pp 31-48",
      SLACcitation   = "%%CITATION = INSPIRE-205976;%%"
}

@article{Ferko:2024zth,
    author = "Ferko, Christian and Kuzenko, Sergei M. and Lechner, Kurt and Sorokin, Dmitri P. and Tartaglino-Mazzucchelli, Gabriele",
    title = "{Interacting chiral form field theories and $ T\overline{T} $-like flows in six and higher dimensions}",
    eprint = "2402.06947",
    archivePrefix = "arXiv",
    primaryClass = "hep-th",
    doi = "10.1007/JHEP05(2024)320",
    journal = "JHEP",
    volume = "05",
    pages = "320",
    year = "2024"
}

@article{Hutomo:2025dfx,
    author = "Hutomo, Jessica and Lechner, Kurt and Sorokin, Dmitri P.",
    title = "{On non-linear chiral 4-form theories in D=10}",
    eprint = "2509.14351",
    archivePrefix = "arXiv",
    primaryClass = "hep-th",
    month = "9",
    year = "2025"
}

@article{Avohou:2019qrl,
    author = "Avohou, Remi C. and Ben Geloun, Joseph and Dub, Nicolas",
    title = "{On the counting of $O(N)$ tensor invariants}",
    eprint = "1907.04668",
    archivePrefix = "arXiv",
    primaryClass = "math-ph",
    doi = "10.4310/ATMP.2020.v24.n4.a1",
    journal = "Adv. Theor. Math. Phys.",
    volume = "24",
    number = "4",
    pages = "821--878",
    year = "2020"
}

@inproceedings{Gaillard:1997rt,
    author = "Gaillard, Mary K. and Zumino, Bruno",
    archivePrefix = "arXiv",
    booktitle = "{Duality and supersymmetric theories. Proceedings, Easter School, Newton Institute, Euroconference, Cambridge, UK, April 7-18, 1997}",
    eprint = "hep-th/9712103",
    month = "12",
    pages = "33--48",
    reportNumber = "LBL-41110, LBNL-41110, UCB-PTH-97-58",
    title = "{Nonlinear electromagnetic selfduality and Legendre transformations}",
    year = "1997"
}

@article{Gaillard:1997zr,
    author = "Gaillard, Mary K. and Zumino, Bruno",
    archivePrefix = "arXiv",
    doi = "10.1007/BFb0105236",
    eprint = "hep-th/9705226",
    journal = "Lect.\ Notes Phys.",
    pages = "121",
    reportNumber = "LBL-40770, LBNL-40770, LBL-40370, UCB-PTH-97-29",
    title = "{Selfduality in nonlinear electromagnetism}",
    volume = "509",
    year = "1998"
}

@article{Schellstede:2016zue,
    author = {Schellstede, Gerold O. and Perlick, Volker and L{\"a}mmerzahl, Claus},
    title = "{On causality in nonlinear vacuum electrodynamics of the Pleba{\'n}ski class}",
    eprint = "1604.02545",
    archivePrefix = "arXiv",
    primaryClass = "gr-qc",
    doi = "10.1002/andp.201600124",
    journal = "Annalen Phys.",
    volume = "528",
    number = "9-10",
    pages = "738--749",
    year = "2016"
}

@article{Russo:2024llm,
    author = "Russo, Jorge G. and Townsend, Paul K.",
    title = "{Causal self-dual electrodynamics}",
    eprint = "2401.06707",
    archivePrefix = "arXiv",
    primaryClass = "hep-th",
    doi = "10.1103/PhysRevD.109.105023",
    journal = "Phys. Rev. D",
    volume = "109",
    number = "10",
    pages = "105023",
    year = "2024"
}

@article{deMelloKoch:2025rkw,
    author = "de Mello Koch, Robert and Jevicki, Antal",
    title = "{Hilbert space of finite N multi-matrix models}",
    eprint = "2508.11986",
    archivePrefix = "arXiv",
    primaryClass = "hep-th",
    doi = "10.1007/JHEP11(2025)145",
    journal = "JHEP",
    volume = "11",
    pages = "145",
    year = "2025"
}

@article{deMelloKoch:2025ngs,
    author = "de Mello Koch, Robert and Jevicki, Antal",
    title = "{Structure of loop space at finite N}",
    eprint = "2503.20097",
    archivePrefix = "arXiv",
    primaryClass = "hep-th",
    doi = "10.1007/JHEP06(2025)011",
    journal = "JHEP",
    volume = "06",
    pages = "011",
    year = "2025"
}

@article{deMelloKoch:2020agz,
    author = "de Mello Koch, Robert and Huang, Jia-Hui and Kim, Minkyoo and Van Zyl, Hendrik J. R.",
    title = "{Emergent Yang-Mills theory}",
    eprint = "2005.02731",
    archivePrefix = "arXiv",
    primaryClass = "hep-th",
    doi = "10.1007/JHEP10(2020)100",
    journal = "JHEP",
    volume = "10",
    pages = "100",
    year = "2020"
}

@article{abo2015eigenconfigurations,
  title={Eigenconfigurations of tensors},
  author={Abo, Hirotachi and Seigal, Anna and Sturmfels, Bernd},
  journal={arXiv preprint arXiv:1505.05729},
  year={2015}
}

@article{Dolan:2007rq,
    author = "Dolan, F. A.",
    title = "{Counting BPS operators in N=4 SYM}",
    eprint = "0704.1038",
    archivePrefix = "arXiv",
    primaryClass = "hep-th",
    reportNumber = "DIAS-STP-07-05",
    doi = "10.1016/j.nuclphysb.2007.07.026",
    journal = "Nucl. Phys. B",
    volume = "790",
    pages = "432--464",
    year = "2008"
}

@article{Mainiero:2019enr,
    author = "Mainiero, Tom",
    title = "{Homological Tools for the Quantum Mechanic}",
    eprint = "1901.02011",
    archivePrefix = "arXiv",
    primaryClass = "hep-th",
    month = "1",
    year = "2019"
}

@Article{ig1,
    title = {The igraph software package for complex network research},
    author = {Gábor Csárdi and Tamás Nepusz},
    journal = {InterJournal},
    volume = {Complex Systems},
    pages = {1695},
    year = {2006},
    url = {https://igraph.org},
  }

@Article{ig2,
    title = {igraph enables fast and robust network analysis across
      programming languages},
    author = {Michael Antonov and Gábor Csárdi and Szabolcs Horvát and
      Kirill Müller and Tamás Nepusz and Daniel Noom and Maëlle Salmon
      and Vincent Traag and Brooke Foucault Welles and Fabio Zanini},
    journal = {arXiv preprint arXiv:2311.10260},
    year = {2023},
    doi = {10.48550/arXiv.2311.10260},
  }

@Manual{ig3,
    title = {{igraph}: Network Analysis and Visualization in R},
    author = {Gábor Csárdi and Tamás Nepusz and Vincent Traag and
      Szabolcs Horvát and Fabio Zanini and Daniel Noom and Kirill
      Müller and David Schoch and Maëlle Salmon},
    year = {2025},
    note = {R package version 2.2.1},
    doi = {10.5281/zenodo.7682609},
    url = {https://CRAN.R-project.org/package=igraph},
  }

@article{Ferko:2025jrf,
    author = "Ferko, Christian and Furuya, Keiichiro",
    title = "{Entanglement cohomology for GHZ and W states}",
    eprint = "2512.19889",
    archivePrefix = "arXiv",
    primaryClass = "hep-th",
    month = "12",
    year = "2025"
}

@article{Ferko:2024swt,
    author = "Ferko, Christian and Iyer, Eashan and Mossayebi, Kasra and Sanfey, Gregor",
    title = "{Hodge theory for entanglement cohomology}",
    eprint = "2410.12529",
    archivePrefix = "arXiv",
    primaryClass = "hep-th",
    doi = "10.1103/PhysRevA.111.032422",
    journal = "Phys. Rev. A",
    volume = "111",
    number = "3",
    pages = "032422",
    year = "2025"
}

@article{Pasti:1995ii,
    author = "Pasti, Paolo and Sorokin, Dmitri P. and Tonin, Mario",
    title = "{Note on manifest Lorentz and general coordinate invariance in duality symmetric models}",
    eprint = "hep-th/9503182",
    archivePrefix = "arXiv",
    reportNumber = "DFPD-95-TH-16",
    doi = "10.1016/0370-2693(95)00463-U",
    journal = "Phys. Lett. B",
    volume = "352",
    pages = "59--63",
    year = "1995"
}

@Article{         harris2020array,
 title         = {Array programming with {NumPy}},
 author        = {Charles R. Harris and K. Jarrod Millman and St{\'{e}}fan J.
                 van der Walt and Ralf Gommers and Pauli Virtanen and David
                 Cournapeau and Eric Wieser and Julian Taylor and Sebastian
                 Berg and Nathaniel J. Smith and Robert Kern and Matti Picus
                 and Stephan Hoyer and Marten H. van Kerkwijk and Matthew
                 Brett and Allan Haldane and Jaime Fern{\'{a}}ndez del
                 R{\'{i}}o and Mark Wiebe and Pearu Peterson and Pierre
                 G{\'{e}}rard-Marchant and Kevin Sheppard and Tyler Reddy and
                 Warren Weckesser and Hameer Abbasi and Christoph Gohlke and
                 Travis E. Oliphant},
 year          = {2020},
 month         = sep,
 journal       = {Nature},
 volume        = {585},
 number        = {7825},
 pages         = {357--362},
 doi           = {10.1038/s41586-020-2649-2},
 publisher     = {Springer Science and Business Media {LLC}},
 url           = {https://doi.org/10.1038/s41586-020-2649-2}
}

@article{Smith2018, doi = {10.21105/joss.00753}, url = {https://doi.org/10.21105/joss.00753}, year = {2018}, publisher = {The Open Journal}, volume = {3}, number = {26}, pages = {753}, author = {Smith, Daniel G. a. and Gray, Johnnie}, title = {opt\_einsum - A Python package for optimizing contraction order for einsum-like expressions}, journal = {Journal of Open Source Software} }

@techreport{hagberg2008exploring,
  title={Exploring network structure, dynamics, and function using NetworkX},
  author={Hagberg, Aric and Swart, Pieter and S Chult, Daniel},
  year={2008},
  institution={Los Alamos National Lab.(LANL), Los Alamos, NM (United States)}
}

@article{Linch:2012zh,
    author = "Linch, III, William D. and Tartaglino-Mazzucchelli, Gabriele",
    title = "{Six-dimensional Supergravity and Projective Superfields}",
    eprint = "1204.4195",
    archivePrefix = "arXiv",
    primaryClass = "hep-th",
    reportNumber = "UUITP-09-12, PP-012-010",
    doi = "10.1007/JHEP08(2012)075",
    journal = "JHEP",
    volume = "08",
    pages = "075",
    year = "2012"
}

@book{travis1999graphical,
  title={Graphical enumeration: A species-theoretic approach},
  author={Travis, Leopold E},
  year={1999},
  publisher={Brandeis University}
}

@article{Kennedy:2025nzm,
    author = "Kennedy, Christian and Tartaglino-Mazzucchelli, Gabriele",
    title = "{Six-dimensional $ \mathcal{N} $ = (2, 0) conformal superspace}",
    eprint = "2506.01630",
    archivePrefix = "arXiv",
    primaryClass = "hep-th",
    doi = "10.1007/JHEP08(2025)215",
    journal = "JHEP",
    volume = "08",
    pages = "215",
    year = "2025"
}

@misc{graphSequence,
  author = "{{OEIS Foundation Inc. (2019)}}",
  title = "{{The On-Line Encyclopedia of Integer Sequences}}",
  howpublished = "{\url{https://oeis.org/A129416}}",
}

@misc{zeoliteOne,
  author = "{{OEIS Foundation Inc. (2019)}}",
  title = "{{The On-Line Encyclopedia of Integer Sequences}}",
  howpublished = "{\url{https://oeis.org/A008005}}",
}

@misc{zeoliteTwo,
  author = "{{OEIS Foundation Inc. (2019)}}",
  title = "{{The On-Line Encyclopedia of Integer Sequences}}",
  howpublished = "{\url{https://oeis.org/A008009}}",
}

@article{Butter:2016qkx,
    author = "Butter, Daniel and Kuzenko, Sergei M. and Novak, Joseph and Theisen, Stefan",
    title = "{Invariants for minimal conformal supergravity in six dimensions}",
    eprint = "1606.02921",
    archivePrefix = "arXiv",
    primaryClass = "hep-th",
    reportNumber = "NIKHEF-2016-026",
    doi = "10.1007/JHEP12(2016)072",
    journal = "JHEP",
    volume = "12",
    pages = "072",
    year = "2016"
}

@article{scikit-learn,
  title={Scikit-learn: Machine Learning in {P}ython},
  author={Pedregosa, F. and Varoquaux, G. and Gramfort, A. and Michel, V.
          and Thirion, B. and Grisel, O. and Blondel, M. and Prettenhofer, P.
          and Weiss, R. and Dubourg, V. and Vanderplas, J. and Passos, A. and
          Cournapeau, D. and Brucher, M. and Perrot, M. and Duchesnay, E.},
  journal={Journal of Machine Learning Research},
  volume={12},
  pages={2825--2830},
  year={2011}
}

@inproceedings{sklearn_api,
  author    = {Lars Buitinck and Gilles Louppe and Mathieu Blondel and
                Fabian Pedregosa and Andreas Mueller and Olivier Grisel and
                Vlad Niculae and Peter Prettenhofer and Alexandre Gramfort
                and Jaques Grobler and Robert Layton and Jake VanderPlas and
                Arnaud Joly and Brian Holt and Ga{\"{e}}l Varoquaux},
  title     = {{API} design for machine learning software: experiences from the scikit-learn
                project},
  booktitle = {ECML PKDD Workshop: Languages for Data Mining and Machine Learning},
  year      = {2013},
  pages = {108--122},
}

@article{Dolan:2008qi,
    author = "Dolan, F. A. and Osborn, H.",
    title = "{Applications of the Superconformal Index for Protected Operators and q-Hypergeometric Identities to N=1 Dual Theories}",
    eprint = "0801.4947",
    archivePrefix = "arXiv",
    primaryClass = "hep-th",
    reportNumber = "DAMTP-08-07, DIAS-STP-08-02, SHEP-08-06",
    doi = "10.1016/j.nuclphysb.2009.01.028",
    journal = "Nucl. Phys. B",
    volume = "818",
    pages = "137--178",
    year = "2009"
}

@article{Hanany:2008sb,
    author = "Hanany, Amihay and Mekareeya, Noppadol and Torri, Giuseppe",
    title = "{The Hilbert Series of Adjoint SQCD}",
    eprint = "0812.2315",
    archivePrefix = "arXiv",
    primaryClass = "hep-th",
    reportNumber = "IMPERIAL-TP-08-AH-11",
    doi = "10.1016/j.nuclphysb.2009.09.016",
    journal = "Nucl. Phys. B",
    volume = "825",
    pages = "52--97",
    year = "2010"
}

@article{Hanany:2007zz,
    author = "Hanany, Amihay",
    editor = "Misra, Aalok",
    title = "{Counting BPS operators in the chiral ring: The plethystic story}",
    doi = "10.1063/1.2803801",
    journal = "AIP Conf. Proc.",
    volume = "939",
    number = "1",
    pages = "165--175",
    year = "2007"
}

@article{Romelsberger:2005eg,
    author = "Romelsberger, Christian",
    title = "{Counting chiral primaries in N = 1, d=4 superconformal field theories}",
    eprint = "hep-th/0510060",
    archivePrefix = "arXiv",
    doi = "10.1016/j.nuclphysb.2006.03.037",
    journal = "Nucl. Phys. B",
    volume = "747",
    pages = "329--353",
    year = "2006"
}

@article{Pouliot:1998yv,
    author = "Pouliot, Philippe",
    title = "{Molien function for duality}",
    eprint = "hep-th/9812015",
    archivePrefix = "arXiv",
    doi = "10.1088/1126-6708/1999/01/021",
    journal = "JHEP",
    volume = "01",
    pages = "021",
    year = "1999"
}

@article{Gripaios_2021,
   title={Lorentz- and permutation-invariants of particles},
   volume={54},
   ISSN={1751-8121},
   url={http://dx.doi.org/10.1088/1751-8121/abe58c},
   DOI={10.1088/1751-8121/abe58c},
   number={15},
   journal={Journal of Physics A: Mathematical and Theoretical},
   publisher={IOP Publishing},
   author={Gripaios, Ben and Haddadin, Ward and Lester, Christopher G},
   year={2021},
   month=mar, pages={155201} }

@misc{gripaios2020lorentzpermutationinvariantsparticles,
      title={Lorentz and permutation invariants of particles II}, 
      author={Ben Gripaios and Ward Haddadin and C. G. Lester},
      year={2020},
      eprint={2007.05746},
      archivePrefix={arXiv},
      primaryClass={hep-th},
      url={https://arxiv.org/abs/2007.05746}, 
}

@article{Lester:2020jrg,
    author = "Lester, Christopher G. and Haddadin, Ward and Gripaios, Ben",
    title = "{Lorentz and permutation invariants of particles III: Constraining nonstandard sources of parity violation}",
    eprint = "2008.05206",
    archivePrefix = "arXiv",
    primaryClass = "hep-ph",
    reportNumber = "CAVENDISH-HEP-20/10",
    doi = "10.1142/S0217751X22500932",
    journal = "Int. J. Mod. Phys. A",
    volume = "37",
    number = "16",
    pages = "2250093",
    year = "2022"
}

@article{Avohou:2024agh,
    author = "Avohou, Remi Cocou and Ben Geloun, Joseph and Toriumi, Reiko",
    title = "{Counting $U(N)^{\otimes r}\otimes O(N)^{\otimes q}$ invariants and tensor model observables}",
    eprint = "2404.16404",
    archivePrefix = "arXiv",
    primaryClass = "hep-th",
    doi = "10.1140/epjc/s10052-024-13091-z",
    journal = "Eur. Phys. J. C",
    volume = "84",
    number = "8",
    pages = "839",
    year = "2024"
}

@article{Chandra:2023afu,
    author = "Chandra, Aditi and Constantin, Andrei and Fraser-Taliente, Cristofero S. and Harvey, Thomas R. and Lukas, Andre",
    title = "{Enumerating Calabi-Yau Manifolds: Placing Bounds on the Number of Diffeomorphism Classes in the Kreuzer-Skarke List}",
    eprint = "2310.05909",
    archivePrefix = "arXiv",
    primaryClass = "hep-th",
    doi = "10.1002/prop.202300264",
    journal = "Fortsch. Phys.",
    volume = "72",
    number = "5",
    pages = "2300264",
    year = "2024"
}

@article{Delporte:2024izt,
    author = "Delporte, Nicolas and Sasakura, Naoki",
    title = "{The edge of random tensor eigenvalues with deviation}",
    eprint = "2405.07731",
    archivePrefix = "arXiv",
    primaryClass = "hep-th",
    doi = "10.1007/JHEP01(2025)071",
    journal = "JHEP",
    volume = "01",
    pages = "071",
    year = "2025"
}

@article{Sasakura:2024awt,
    author = "Sasakura, Naoki",
    title = "{Signed Eigenvalue/vector Distribution of Complex Order-Three Random Tensor}",
    eprint = "2404.03385",
    archivePrefix = "arXiv",
    primaryClass = "hep-th",
    reportNumber = "YITP-24-36",
    doi = "10.1093/ptep/ptae062",
    journal = "PTEP",
    volume = "2024",
    number = "5",
    pages = "053A04",
    year = "2024"
}

@article{QI20051302,
title = {Eigenvalues of a real supersymmetric tensor},
journal = {Journal of Symbolic Computation},
volume = {40},
number = {6},
pages = {1302-1324},
year = {2005},
issn = {0747-7171},
doi = {https://doi.org/10.1016/j.jsc.2005.05.007},
url = {https://www.sciencedirect.com/science/article/pii/S0747717105000817},
author = {Liqun Qi}
}

@ARTICLE{2010arXiv1004.4953C,
       author = {{Cartwright}, Dustin and {Sturmfels}, Bernd},
        title = "{The Number of Eigenvalues of a Tensor}",
      journal = {arXiv e-prints},
         year = 2010,
        month = apr,
          eid = {arXiv:1004.4953},
        pages = {arXiv:1004.4953},
          doi = {10.48550/arXiv.1004.4953},
archivePrefix = {arXiv},
       eprint = {1004.4953},
 primaryClass = {math.NA},
       adsurl = {https://ui.adsabs.harvard.edu/abs/2010arXiv1004.4953C}
}

@article{Cremonini:2022cdm,
    author = "Cremonini, C. A. and Grassi, P. A. and Noris, R. and Ravera, L.",
    title = "{Supergravities and branes from Hilbert-Poincar{\'e} series}",
    eprint = "2211.10454",
    archivePrefix = "arXiv",
    primaryClass = "hep-th",
    doi = "10.1007/JHEP12(2023)088",
    journal = "JHEP",
    volume = "12",
    pages = "088",
    year = "2023"
}

@inproceedings{Qi2018TensorEA,
  title={Tensor Eigenvalues and Their Applications},
  author={Liqun Qi and Yannan Chen and Haibin Chen},
  year={2018},
  url={https://api.semanticscholar.org/CorpusID:126083359}
}

@article{deMelloKoch:2025qeq,
    author = "de Mello Koch, Robert and Kim, Minkyoo and Van Zyl, Hendrik J. R.",
    title = "{From Symmetry to Structure: Gauge-Invariant Operators in Multi-Matrix Quantum Mechanics}",
    eprint = "2507.01219",
    archivePrefix = "arXiv",
    primaryClass = "hep-th",
    month = "7",
    year = "2025"
}

@article{PROCESI198763,
title = {A formal inverse to the Cayley-Hamilton theorem},
journal = {Journal of Algebra},
volume = {107},
number = {1},
pages = {63-74},
year = {1987},
issn = {0021-8693},
doi = {https://doi.org/10.1016/0021-8693(87)90073-1},
url = {https://www.sciencedirect.com/science/article/pii/0021869387900731},
author = {Claudio Procesi}
}

@book{nagata1962local,
  title={Local Rings},
  author={Nagata, M.},
  isbn={9780470628652},
  lccn={62017459},
  series={Interscience tracts in pure and applied mathematics},
  url={https://books.google.com/books?id=2FngCHSdpnYC},
  year={1962},
  publisher={Interscience Publishers}
}

@article{Cohen1946CompleteLocalRings,
  author  = {Cohen, I. S.},
  title   = {On the Structure and Ideal Theory of Complete Local Rings},
  journal = {Transactions of the American Mathematical Society},
  volume  = {59},
  number  = {1},
  year    = {1946},
  pages   = {54--106},
  doi     = {10.1090/S0002-9947-1946-0016094-3},
  url     = {https://www.ams.org/tran/1946-059-01/S0002-9947-1946-0016094-3/}
}

@book{Macaulay1916ModularSystems,
  author    = {Macaulay, Francis Sowerby},
  title     = {The Algebraic Theory of Modular Systems},
  year      = {1916},
  publisher = {Cambridge University Press},
  address   = {Cambridge},
  series    = {Cambridge Tracts in Mathematics and Mathematical Physics},
  number    = {19},
  pages     = {112},
  url       = {https://archive.org/details/algebraictheoryo00macauoft}
}

@book{bruns1998cohen,
  title={Cohen-Macaulay Rings},
  author={Bruns, W. and Herzog, H.J.},
  isbn={9780521566742},
  lccn={gb98053830},
  series={Cambridge Studies in Advanced Mathematics},
  url={https://books.google.com/books?id=LF6CbQk9uScC},
  year={1998},
  publisher={Cambridge University Press}
}

@book{derksen2002computational,
  title={Computational Invariant Theory},
  author={Derksen, H. and Kemper, G.},
  isbn={9783540434764},
  lccn={02021729},
  series={Encyclopaedia of Mathematical Sciences},
  url={https://books.google.com/books?id=sHQ5E_t76S8C},
  year={2002},
  publisher={Springer Berlin Heidelberg}
}

@article{Conti:2018jho,
      author         = "Conti, Riccardo and Iannella, Leonardo and Negro, Stefano
                        and Tateo, Roberto",
      title          = "{Generalised Born-Infeld models, Lax operators and the $
                        \mathrm{T}\overline{\mathrm{T}} $ perturbation}",
      journal        = "JHEP",
      volume         = "11",
      year           = "2018",
      pages          = "007",
      doi            = "10.1007/JHEP11(2018)007",
      eprint         = "1806.11515",
      archivePrefix  = "arXiv",
      primaryClass   = "hep-th",
      SLACcitation   = "%%CITATION = ARXIV:1806.11515;%%"
}

@article{Cavaglia:2016oda,
      author         = "Cavagli\`a, Andrea and Negro, Stefano and Sz\'ecs\'enyi, Istv\'an M. and Tateo, Robertoo",
      title          = "{$T \bar{T}$-deformed 2D Quantum Field Theories}",
      journal        = "JHEP",
      volume         = "10",
      year           = "2016",
      pages          = "112",
      doi            = "10.1007/JHEP10(2016)112",
      eprint         = "1608.05534",
      archivePrefix  = "arXiv",
      primaryClass   = "hep-th",
      SLACcitation   = "%%CITATION = ARXIV:1608.05534;%%"
}

@article{Zamolodchikov:2004ce,
      author         = "Zamolodchikov, Alexander B.",
      title          = "{Expectation value of composite field T anti-T in
                        two-dimensional quantum field theory}",
      year           = "2004",
      eprint         = "hep-th/0401146",
      archivePrefix  = "arXiv",
      primaryClass   = "hep-th",
      reportNumber   = "BONN-TH-2004-02",
      SLACcitation   = "%%CITATION = HEP-TH/0401146;%%"
}

@article{Smirnov:2016lqw,
      author         = "Smirnov, F. A. and Zamolodchikov, A. B.",
      title          = "{On space of integrable quantum field theories}",
      journal        = "Nucl. Phys.",
      volume         = "B915",
      year           = "2017",
      pages          = "363-383",
      doi            = "10.1016/j.nuclphysb.2016.12.014",
      eprint         = "1608.05499",
      archivePrefix  = "arXiv",
      primaryClass   = "hep-th",
      SLACcitation   = "%%CITATION = ARXIV:1608.05499;%%"
}

@article{Bonelli:2018kik,
	Archiveprefix = {arXiv},
	Author = {Bonelli, Giulio and Doroud, Nima and Zhu, Mengqi},
	Doi = {10.1007/JHEP06(2018)149},
	Eprint = {1804.10967},
	Journal = {JHEP},
	Pages = {149},
	Primaryclass = {hep-th},
	Slaccitation = {%%CITATION = ARXIV:1804.10967;%%},
	Title = {{$T \bar{T}$-deformations in closed form}},
	Volume = {06},
	Year = {2018}}

@article{Ebert:2024zwv,
    author = "Ebert, Stephen and Ferko, Christian and Martin, Cian Luke and Tartaglino-Mazzucchelli, Gabriele",
    title = "{Flows in the space of interacting chiral boson theories}",
    eprint = "2403.18242",
    archivePrefix = "arXiv",
    primaryClass = "hep-th",
    doi = "10.1103/PhysRevD.110.046005",
    journal = "Phys. Rev. D",
    volume = "110",
    number = "4",
    pages = "046005",
    year = "2024"
}

@article{Avetisyan:2022zza,
    author = "Avetisyan, Zhirayr and Evnin, Oleg and Mkrtchyan, Karapet",
    title = "{Nonlinear (chiral) p-form electrodynamics}",
    eprint = "2205.02522",
    archivePrefix = "arXiv",
    primaryClass = "hep-th",
    reportNumber = "Imperial-TP-KM-2022-1",
    doi = "10.1007/JHEP08(2022)112",
    journal = "JHEP",
    volume = "08",
    pages = "112",
    year = "2022"
}

@article{Ferko:2023sps,
    author = "Ferko, Christian and Hu, Yangrui and Huang, Zejun and Koutrolikos, Konstantinos and Tartaglino-Mazzucchelli, Gabriele",
    title = "{$T \overline{T}$-like flows and $3d$ nonlinear supersymmetry}",
    eprint = "2302.10410",
    archivePrefix = "arXiv",
    primaryClass = "hep-th",
    doi = "10.21468/SciPostPhys.16.1.038",
    journal = "SciPost Phys.",
    volume = "16",
    number = "1",
    pages = "038",
    year = "2024"
}

@article{Ferko:2024yua,
    author = "Ferko, Christian and Hou, Jue and Morone, Tommaso and Tartaglino-Mazzucchelli, Gabriele and Tateo, Roberto",
    title = "{TT-like Flows of Yang-Mills Theories}",
    eprint = "2409.18740",
    archivePrefix = "arXiv",
    primaryClass = "hep-th",
    doi = "10.1103/PhysRevLett.134.101603",
    journal = "Phys. Rev. Lett.",
    volume = "134",
    number = "10",
    pages = "101603",
    year = "2025"
}

@article{Ferko:2023wyi,
    author = "Ferko, Christian and Kuzenko, Sergei M. and Smith, Liam and Tartaglino-Mazzucchelli, Gabriele",
    title = "{Duality-invariant nonlinear electrodynamics and stress tensor flows}",
    eprint = "2309.04253",
    archivePrefix = "arXiv",
    primaryClass = "hep-th",
    doi = "10.1103/PhysRevD.108.106021",
    journal = "Phys. Rev. D",
    volume = "108",
    number = "10",
    pages = "106021",
    year = "2023"
}

@article{Ferko:2023ruw,
    author = "Ferko, Christian and Smith, Liam and Tartaglino-Mazzucchelli, Gabriele",
    title = "{Stress Tensor flows, birefringence in non-linear electrodynamics and supersymmetry}",
    eprint = "2301.10411",
    archivePrefix = "arXiv",
    primaryClass = "hep-th",
    doi = "10.21468/SciPostPhys.15.5.198",
    journal = "SciPost Phys.",
    volume = "15",
    number = "5",
    pages = "198",
    year = "2023"
}

@article{Ferko:2022iru,
    author = "Ferko, Christian and Smith, Liam and Tartaglino-Mazzucchelli, Gabriele",
    title = "{On Current-Squared Flows and ModMax Theories}",
    eprint = "2203.01085",
    archivePrefix = "arXiv",
    primaryClass = "hep-th",
    doi = "10.21468/SciPostPhys.13.2.012",
    journal = "SciPost Phys.",
    volume = "13",
    number = "2",
    pages = "012",
    year = "2022"
}

@article{Ferko:2022dpg,
    author = "Ferko, Christian and Sethi, Savdeep",
    title = "{Sequential flows by irrelevant operators}",
    eprint = "2206.04787",
    archivePrefix = "arXiv",
    primaryClass = "hep-th",
    reportNumber = "EFI-22-05",
    doi = "10.21468/SciPostPhys.14.5.098",
    journal = "SciPost Phys.",
    volume = "14",
    number = "5",
    pages = "098",
    year = "2023"
}
\end{document}